\documentclass[10pt,aps,prd,amsmath,amssymb,superscriptaddress,twocolumn,nofootinbib,showpacs,preprintnumbers,amsfonts, notitlepage]{revtex4-1}
\usepackage{XYZ}

\begin{document}

\title{\boldmath \XYZ spectroscopy at electron-hadron facilities II:\\Semi-inclusive processes with pion exchange}

\author{D.~Winney}
\email{dwinney@scnu.edu.cn}
\affiliation{\scnuIQM}
\affiliation{\scnuJLQM}

\author{A.~Pilloni}
\email{alessandro.pilloni@unime.it}
\affiliation{\messina}
\affiliation{\catania}

\author{V.~Mathieu}
\email{vmathieu@ub.edu}
\affiliation{\ub}

\author{A.~N.~Hiller Blin}
\affiliation{\ur}
\affiliation{\ut}

\author{M.~Albaladejo}
\affiliation{\ific}

\author{W.~A.~Smith}
\affiliation{\ceem}
\affiliation{\indiana}

\author{A.~Szczepaniak}
\affiliation{\ceem}
\affiliation{\indiana}
\affiliation{\jlab}

\collaboration{Joint Physics Analysis Center}


\newcommand{\catania}{INFN Sezione di Catania, I-95123 Catania, Italy}
\newcommand{\ceem}{Center for  Exploration  of  Energy  and  Matter, Indiana  University, Bloomington,  IN  47403,  USA}
\newcommand{\cern}{CERN, 1211 Geneva 23, Switzerland}
\newcommand{\icn}{Instituto de Ciencias Nucleares,
Universidad Nacional Aut\'onoma de M\'exico, Ciudad de M\'exico 04510, Mexico}
\newcommand{\icsup}{Pedagogical University of Krakow, 30-084 Krak\'ow, Poland}
\newcommand{\ific}{Instituto de F\'isica Corpuscular (IFIC), Centro Mixto CSIC-Universidad de Valencia, E-46071 Valencia, Spain}
\newcommand{\indiana}{Department of Physics,
Indiana  University, Bloomington,  IN  47405,  USA}
\newcommand{\jlab}{Theory Center, Thomas  Jefferson  National  Accelerator  Facility, Newport  News,  VA  23606,  USA}
\newcommand{\ut}{Institute for Theoretical Physics, T\"ubingen University, Auf der Morgenstelle 14, 72076 T\"ubingen, Germany}
\newcommand{\ur}{Institute for Theoretical Physics, Regensburg University, D-93040 Regensburg, Germany}
\newcommand{\lanl}{Theoretical Division, Los Alamos National Laboratory, Los Alamos, NM 87545, USA}
\newcommand{\lmu}{Ludwig-Maximilian University of Munich, Germany}
\newcommand{\messina}{Dipartimento di Scienze Matematiche e Informatiche, Scienze Fisiche e Scienze della Terra,
Universit\`a degli Studi di Messina, I-98122 Messina, Italy}
\newcommand{\odu}{Department of Physics, Old Dominion University, Norfolk, VA 23529, USA}
\newcommand{\origins}{ORIGINS Excellence Cluster, 80939 Munich, Germany}
\newcommand{\scnuIQM}{Guangdong Provincial Key Laboratory of Nuclear Science, Institute of Quantum Matter, South China Normal University, Guangzhou 510006, China}
\newcommand{\scnuJLQM}{Guangdong-Hong Kong Joint Laboratory of Quantum Matter, Southern Nuclear Science Computing Center, South China Normal University, Guangzhou 510006, China}
\newcommand{\ub}{Departament de F\'isica Qu\`antica i Astrof\'isica and Institut de Ci\`encies del Cosmos, Universitat de Barcelona, E-08028, Spain}
\newcommand{\ucm}{Departamento de F\'isica Te\'orica, Universidad Complutense de Madrid and IPARCOS, E-28040 Madrid, Spain}
\newcommand{\uned}{Departamento de F\'isica Interdisciplinar, Universidad Nacional de Educaci\'on a Distancia (UNED), Madrid E-28040, Spain}
\newcommand{\wm}{College of William \& Mary, Williamsburg, VA 23187, USA}

\begin{abstract}
Semi-inclusive processes are very promising to investigate \XYZ hadrons at the next generation of electron-hadron facilities, because they generally boast higher cross sections. We extend our formalism of exclusive photoproduction to semi-inclusive final states. The inclusive production cross sections for charged axial-vector $Z$ states from pion exchange are predicted.  We isolate the contribution of $\Delta$ resonances at small missing mass. Production near threshold is shown to be enhanced roughly by a factor of two compared to the exclusive reaction. We benchmark the model with data of semi-inclusive $b_1^\pm$ production.
\end{abstract}

\preprint{JLAB-THY-22-3719}
\maketitle

\section{Introduction}
\label{sec:intro}
Appearance of the exotic \XYZ states in the spectrum of heavy quarkonia is widely recognized as one of the most intriguing puzzles
  with potentially high impact on our understanding of QCD~\cite{Esposito:2016noz,*Guo:2017jvc,*Olsen:2017bmm,*Wallbott:2019dng,*Brambilla:2019esw,JPAC:2021rxu}. 
 Most of these states have been observed only in specific  
  channels,  most notably in heavy hadron decays and via direct production in $e^+e^-$ collisions~\cite{BESIII:2013ris,*BESIII:2016bnd,*Belle:2003nnu}. Exploring alternative production processes, such as electro- or photoproduction can 
    provide complementary information on the  nature of these states, while probing if they are real resonances or mere kinematic effects~\cite{Guo:2019twa}.
  
In a previous paper~\cite{Albaladejo:2020tzt}, we calculated production rates for several of these states in exclusive photo- and electroproduction, at energies that have been proposed, both for the future Electron Ion Collider  (EIC)~\cite{AbdulKhalek:2021gbh} and a new facility that could take advantage of an  energy upgrade of the CEBAF accelerator~\cite{Arrington:2021alx}. 
While exclusive reactions benefit from constrained kinematics, complementary information can be obtained from inclusive reactions. For example, there is ample literature on inclusive   $X(3872)$ production, \eg in heavy-ion collisions and how it is relevant in  unraveling its composition~\cite{Bignamini:2009sk,*Bignamini:2009fn,Artoisenet:2009wk,*Artoisenet:2010uu,Albaladejo:2017blx,*Esposito:2017qef,Esposito:2020ywk,*Braaten:2020iqw,AbdulKhalek:2021gbh}. 
Hard events can be studied with perturbative QCD and effective field theories, and one can perform global fits of the long-distance matrix elements from electron-hadron and hadron-hadron collisions, to be compared with model predictions~\cite{Artoisenet:2009wk,*Artoisenet:2010uu}. Soft processes are dominated by specific kinematic configurations. 
Compared to exclusive production, inclusive reactions benefit from larger  cross sections, and often rely  less on  model assumptions. 
  
In this paper, we focus on inclusive production of states  that can process via 
 one pion exchange. Modulo final state interactions, pion exchange is a rather well tested hypothesis and given its proximity to the physical threshold  it usually  results in large cross sections. 
We test our model by comparing with data on the $b_1$ photoproduction. We find a good agreement with the cross section  measured by  the OmegaPhoton collaboration~\cite{OmegaPhoton:1984ols}.  As in our previous work~\cite{Albaladejo:2020tzt},  in  predicting  the photoproduction cross section for  the \Zs states  we  rely on the  measured branching fractions and infer other  properties from  the well-established quarkonium phenomenology. This makes  our predictions as agnostic as possible as far as the nature of these states. 

 The paper is organized as follows.  The  following \cref{sec:formalism} outlines the formalism for  single meson semi-inclusive production. It includes 
discussion of the virtual pion-nucleon cross section, which we study in regimes of both small and large missing mass.
\Cref{sec:numerics} contains numerical results for the inclusive cross sections of axial-vector mesons, in particular the $b_1(1250)$ and \Zs states. Finally, concluding remarks and summary of our results are given in  \cref{sec:conclusions}. For reference we provide a summary of kinematic expressions relevant to inclusive processes in \cref{app:kinematics} as well as formulae connecting SAID partial-waves to the total pion-nucleon cross section in \cref{app:SAID}.

\section{Formalism}
\label{sec:formalism}
    \begin{figure}[t]
        \centering
        \includegraphics[width=.5\columnwidth]{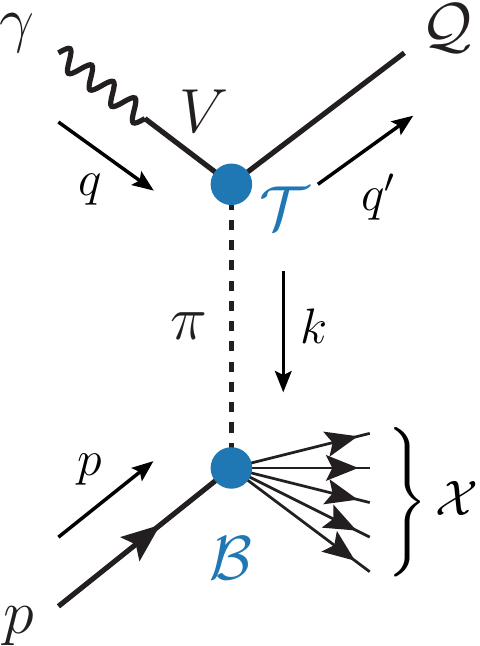}
        \caption{Semi-inclusive photoproduction of an axial-vector \mQ via
pion exchange in the $t$-channel. The bottom vertex $\mB$ is generalized to consider the production of arbitrary final state $\mN$. 
}
        \label{fig:diagram}
    \end{figure}

We consider the process $\gamma p \to \mQ^\pm\, \mN$, where $\mQ$ is an axial-vector (quarkoniumlike) meson with mass $m_\mQ$, and $\mN$ collectively refers to unobserved particles with total invariant mass \M, \aka ``missing mass''. We do not consider production of neutral $\mQ^0$ as it contains  Pomeron exchange which, 
 given the limited  available data for the $Z$ states, will depend more on model assumption. We note that $\mN$ has  baryon quantum numbers and a minimum mass $\Mmin$ which, for $\mQ^+$, is the nucleon mass and corresponds to the exclusive,   $\gamma p \to \mQ^+ n$ reaction. Since this process was already studied in Ref.~\cite{Albaladejo:2020tzt}, in the following we fix the minimum mass to be equal to the first inelastic threshold, \ie  $\Mmin=m_{\pi^0}+m_n$ 
(and for the $\mQ^-$, $\Mmin=m_{\pi^+}+m_p$). The process is represented schematically in \cref{fig:diagram}, and the relevant kinematics is  summarized in \cref{app:kinematics}.

\subsection{Generalized optical theorem}
\begin{figure*}
        \centering
        \includegraphics[width=0.8\textwidth]{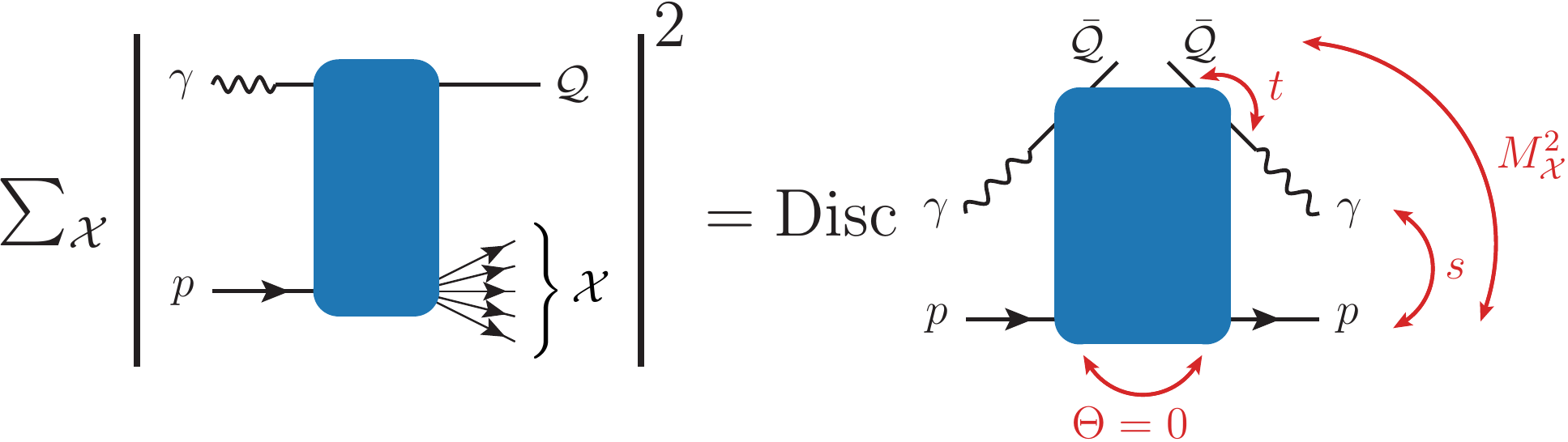}
        \caption{Diagrammatic representation of the generalized optical theorem. The semi-inclusive amplitude squared summed over all possible final states is related to the discontinuity of the $3\to3$ forward elastic scattering amplitude.}
        \label{fig:Mueller}
    \end{figure*}

The extension of the  single-particle exchange mechanism of the exclusive reactions to semi-inclusive  production is given by the generalized optical theorem~\cite{Mueller:1970fa}.
A sketch of the derivation is provided below 
for illustration purposes and for a more detailed discussion we refer to~\cite{Collins:1977jy}.
The Lorentz-invariant differential cross section for the reaction 
$\gamma p \to \mQ^\pm\, \mN$, is given by: 
\begin{widetext}
\begin{equation}
    \label{eq:sumoverX}
 E_\mQ \frac{d^3\sigma}{d^3 q_f} = \frac{1}{16\pi^3}\frac{1}{4E_\gamma \sqrt{s}} \, \frac{1}{4}\sum_{\{\lambda\}}  \, \sum_{\mN} \int \prod_n \frac{d^3p_n }{(2\pi)^3\,2E_n} \, \left|A^{\gamma N \to \mQ\mN}_{\{\lambda\}}\right|^2 (2\pi)^4  \, \delta^4\!\left(q+p - q' - \sum_n p_n \right)~,
\end{equation}
where $\{\lambda\} = \lambda_\gamma, \lambda_N, \lambda_\mQ$ collectively denotes particle helicities, and the sum over $\mN$ runs over all possible final states containing $n$ unobserved particles. By analytical continuation,  crossing symmetry relates the amplitude for $\gamma N \to \mQ \mN$ to that of the reaction   $\gamma N \bar{\mQ} \to \mN$.  By summing over all intermediate states $\mN$, using unitarity for a  $3\to3$ amplitude one can express  \cref{eq:sumoverX} in terms of the discontinuity of the forward elastic $3\to3$ amplitude across the $\M^2$ cut:\footnote{We note that the discontinuity of a $3\to 3$ amplitude depends on eight kinematic variables.  What enters the generalized optical theorem of \cref{eq:disc} is the forward amplitude,  with the final state particles having the same momenta as the initial state particles. In this kinematics there are only three independent variables, denoted, by $s$, $t$, and $\M^2$ as shown in Fig.~\ref{fig:Mueller}. }
    \begin{align}
         \Disc \, A_{\{\lambda\}}^{\gamma N \bar \mQ} &= \frac{1}{2i} \,  \left[ A_{\{\lambda\}}^{\gamma N \bar \mQ}(s, t, \M^2 + \ieps) - A_{\{\lambda\}}^{\gamma N \bar \mQ}(s, t, \M^2 - \ieps)\right] 
        \nonumber \\
        &= \frac{1}{2}\sum_\mN \, \int \prod_n \frac{d^3p_n }{(2\pi)^3\,2E_n} \, \left|A^{\gamma N \to \mQ\mN}_{\{\lambda\}}\right|^2 (2\pi)^4  \, \delta^4\!\left(q+p - q' - \sum_n p_n \right)~.\label{eq:disc}
    \end{align}
\end{widetext}
Comparing with Eq.~\eqref{eq:sumoverX}  one can write, 
\begin{align}\label{eq:xsecFormal}
    E_\mQ \, \frac{d^3\sigma}{d^3q_f} &= \frac{1}{16\pi^3}\frac{1}{2 E_\gamma \sqrt{s}} \,  \frac{1}{4}\sum_{\{\lambda\}} \Disc A^{\gamma N \bar \mQ}_{\lambda_\gamma \lambda_N \lambda_\mQ},
\end{align}
or, in term of Mandelstam variables,
\begin{align}
\frac{\diff^2\sigma}{\diff t \, \diff \M^2} &= \frac{1}{16\pi^2}\frac{1}{4 E^2_\gamma \,s} \,  \frac{1}{4}\sum_{\{\lambda\}} \Disc A^{\gamma N \bar \mQ}_{\lambda_\gamma \lambda_N \lambda_\mQ}.
\end{align}
The semi-inclusive cross section is therefore given by the discontinuity of the $3\to3$ amplitude containing both $t$ and $\M^2$ dependence as depicted in \cref{fig:Mueller}. 
As a check on normalization 
 in \cref{eq:xsecFormal} it is usefully to derive the exclusive formulae from it. Writing explicitly the one-body phase space in \cref{eq:disc}, one obtains,
\begin{align}
    \label{eq:3b_disc}
    \Disc A^{\gamma N \bar \mQ}_{\lambda_\gamma \lambda_N \lambda_\mQ} &= \sum_{\lambda'_N} \left| A^{\gamma N\to \mQ \, N' }_{\lambda_\gamma \lambda_N \lambda_\mQ \lambda'_N} \right|^2 \pi \delta(\M^2-m^2_N),
\end{align}
whence
\begin{align} \nonumber
    \frac{\diff \sigma}{\diff t} & = \int \diff \M^2 \, \frac{\diff^2\sigma}{\diff t \diff\M} 
    \\
    & = \frac{1}{16\pi} \frac{1}{(2 E_\gamma\sqrt{s})^2} \, \frac{1}{4} \sum_{\{\lambda\}}  \left| A^{\gamma N\to \mQ \, N' }_{\lambda_\gamma \lambda_N \lambda_\mQ \lambda'_N} \right|^2~,
\end{align}
which agrees with the  standard expression for the 
  exclusive differential cross
     section~\cite{Irving:1977ea}.
 Furthermore, using 
   the $t$-channel pion exchange model for exclusive  $Z$ production previously considered in   work~\cite{Albaladejo:2020tzt} into \cref{eq:3b_disc}, one obtains, 
    \begin{align}
        \label{TPB}
        \Disc &A^{\gamma N \bar \mQ}_{\lambda_\gamma \lambda_N \lambda_\mQ} = \left|\mT_{\lambda_\gamma\lambda_\mQ} \, \mathcal{P}_\pi \right|^2 
        \nonumber \\
        &\qquad\times \sum_{\lambda^\prime_N}|\mB_{\lambda_N \lambda^\prime_{N}}|^2 \, \pi \, \delta(\M^2 - m^2_N).
    \end{align}
Here, $\mT$ and $\mB$ are the $\pi\gamma\mQ$ and $\pi NN$ vertex functions respectively and $\mathcal{P}_\pi$ is the pion propagator. 

The second line is identified  with  the nucleon pole contribution to the elastic scattering of an off-shell pion ($\pi^*$) off the nucleon.
We may thus write, 
    \begin{equation}
        \label{eq:3b_to_Born}
        \Disc A^{\gamma N \bar \mQ}_{\lambda_\gamma \lambda_N \lambda_\mQ} = \left|\mT_{\lambda_\gamma\lambda_\mQ} \, \mathcal{P}_\pi \right|^2 
        \Disc A^{\pi^* N}_{\lambda_N}~,
    \end{equation}
which shows how the $\M^2$ dependence 
 of the exclusive amplitude generalizes to the inclusive case through the forward (virtual) $\pi N$ scattering amplitude. 
Inserting \cref{eq:3b_to_Born} into \cref{eq:xsecFormal} gives
    \begin{align}
        &\dtsig = \frac{K}{16\pi^3} \, \frac{1}{2} \sum_{\lambda_\gamma \lambda_\mQ}\left|\mT_{\lambda_\gamma\lambda_\mQ} \, \mathcal{P}_\pi \right|^2   \, \sigmatotst~,
    \end{align}
where we also use the optical theorem to express the forward $\pi N$ elastic amplitude in terms of the total cross section, 
    \begin{equation}
        \label{eq:2-2_optical}
        \frac{1}{2} \sum_{\lambda_N}\Im A_{\lambda_N}^{\pi^* N} = \lambda^{1/2}( \M^2, t, m_N^2) \, \sigmatotst~,
    \end{equation}
with $\sigmatotst \equiv \sigmatotst(t, \M^2)$ and 
the  flux factor 
    \begin{equation}
        K \equiv K(s,t,\M^2) = \frac{\lambda^{1/2}( \M^2, t, m_N^2)}{2E_\gamma \sqrt{s}}~.
    \end{equation}

In the approach of~\cite{Albaladejo:2020tzt}, the top vertex is approximated by an effective $\gamma \mQ \pi$ Lagrangian and thus  
in the $t$-channel frame the spin-flip interactions between the photon and $\mQ$ vanish. The sum over helicities in the top vertex then reduces to: 
    \begin{align}
        \label{eq:dtsigma_gen}
        \dtsig = \frac{K}{16\pi^3} \,
         \left|T_\pi(t) \, \mathcal{P}_\pi \right|^2   \, \sigmatotst
       ~,
    \end{align}
where the residue function $T_\pi(t)$ is related to a dimensionless $\pi\gamma\mQ$ coupling constant:
    \begin{align}
        \label{eq:T_pi}
        T_\pi(t) =  g_{\gamma \mQ\pi} \, \frac{\lambda^{1/2}(t, 0, m_\mQ^2)}{2 \, m_\mQ} \; e^{t^\prime / \Lambda_\pi^2}
       ~.
    \end{align}
Here we also include the exponential form factor with
pion cutoff, $\Lambda_\pi = 900$ MeV, to account for the observed momentum-transfer dependence of the OPE cross sections.
For the ease of comparison with the exclusive kinematics and to avoid any  spurious dependence on $\M$, 
 we fix $t^\prime$ to the value it takes for the exclusive reaction, $t^\prime \equiv t - t_\text{min}(s, \M = m_N)$.  
 
With an appropriately chosen pion propagator and parameterization for the (off-shell) $\pi N$ cross section, \cref{eq:dtsigma_gen} becomes the semi-inclusive generalization of the pion-exchange model. The form of the propagator depends on the energy range of interest for the production reaction. For instance, a Feynman diagram-inspired model of fixed-spin pion exchange with a scalar propagator
    \begin{equation}
        \label{eq:mP_FS}
        \mathcal{P}_\pi = \frac{1}{m_\pi^2 - t} ~,
    \end{equation}
is expected to be reliable at energies near threshold. 
In the high energy limit, to leading order  in $s$ the amplitude is given by  Reggeized pion exchanges. In particular, we focus on the so-called triple Regge region, where $s \gg \M \gg |t|$ in which the  amplitude is simply obtained by replacing the pion propagator with
    \begin{align}
        \label{eq:mP_R}
        \mP_\pi
        \to 
        \alpha^\prime \, \xi(t) \, 
        \Gamma(-\alpha(t)) \left(\frac{s}{\M^2}\right)^{\alpha(t)}~,
    \end{align}
in terms of a Regge pole~\cite{Collins:1977jy}.
The signature factor $\xi$ is given by 
    \begin{equation}
        \xi(t) = \frac{1}{2}\left[1+ \tau e^{-i\pi\alpha(t)}\right]~,
    \end{equation}
with $\tau_\pi = +1$ and pion trajectory 
    \begin{equation}
        \alpha_\pi(t) = \alpha_\pi^\prime (t - m_\pi^2) 
        \quad \text{with} \quad 
        \alpha_\pi^\prime = 0.7 \gevsq~.
    \end{equation}
Here we note that compared to the usual Regge pole form, $\M^2$ appears in the denominator of \cref{eq:mP_R} instead of the constant mass scale related to the masses of particles involved, $s_0$. This is because $s\gg \M^2 \gg s_0 \gtrsim |t|$, thus $\cos\theta_t \to s / \M^2$, with $\M^2$ setting the dimensional scale. 

Examining the asymptotic behavior of \cref{eq:mP_R}, we see that at very large $t$ (outside the validity range of the model), the $\Gamma$ function  grows faster than exponentially, exceeding the suppression from the form factor:
    \begin{equation}
        \label{eq:mP_R_asym}
    e^{t^\prime / \Lambda^2} \, \Gamma(-\alpha(t)) \to  e^{t \, \left[\Lambda^{-2} + \alpha^\prime - \alpha^\prime \,\log(\alpha^\prime \, |t|)\right] }
   ~.
    \end{equation}
Thus, in order to avoid unphysical contributions when integrating over the whole semi-inclusive phase-space, in the numerical studies below we impose a cutoff
   \begin{equation}
        |t_\text{cut}| \lesssim \frac{1}{\alpha^\prime} \, e^{1/\Lambda^{2} \,\alpha^\prime} \sim 8.3\gevsq~,
    \end{equation}
 such that the amplitude is exponentially dampened in the entire $t$ range considered. 

\subsection{The  $\pi^* N$ total cross section}
\label{sec:sigmatot}
The aspect of the model in  \cref{eq:dtsigma_gen} not present in the previous analysis of exclusive reactions is the generalized ``bottom vertex" which is given by the total $\pi^* N$ cross section.  Since virtual pion-nucleon scattering cross sections are not known we will relate this as much as possible to the usual on-shell pion-nucleon scattering process. 
Additionally, as before, it is important to consider the different kinematic regimes of the variable $\M^2$ since the near-threshold $\pi N$ spectrum is dominated by nucleon resonances which may dramatically affect the inclusive production. In fact, the resonance region shows vastly different behavior between the  different isospin states which means it is now important to clearly denote the charge of the  channel considered. We fix the initial nucleon to a proton target such that $\mQ^\pm$ production involves $\pi^{\mp}p$ scattering in the bottom vertex.

To this end we use the formalism of Ref.~\cite{Mathieu:2015gxa}, which 
 describes the  total $\pi^{\pm} p$ cross section at all energies. 
At small $\M^2$ the cross section is dominated by the $\Delta$ and $N^*$ resonances. Because the intrinsic properties of these resonances are not the focus of our study, a sophisticated analytic model is not required. Instead, we  interpolate the SAID partial wave amplitudes based on the $T$-matrix analysis of the GW-PWA group~\cite{Workman:2012hx}. Partial waves for both parities and $s$-channel isospin projections
are given up to orbital angular momentum $L = 7$. These may be related to $t$-channel isospin amplitudes by considering isospin crossing combinations as in Ref.~\cite{Mathieu:2015gxa}. We provide a brief summary of these relations in \cref{app:SAID}.
Importantly, the  decomposition is in the form
    \begin{equation}
        \label{eq:sigmaL_pip}
        \sigma^{\pi^\pm p}_\text{tot}(\M^2) = \sum_L \sigma_L^{\pi^\pm p}(\M^2)~,
    \end{equation}
where the right-hand side is calculated from the SAID partial waves at fixed $L$.

    \begin{figure}
        \centering
        \includegraphics[width=\columnwidth]{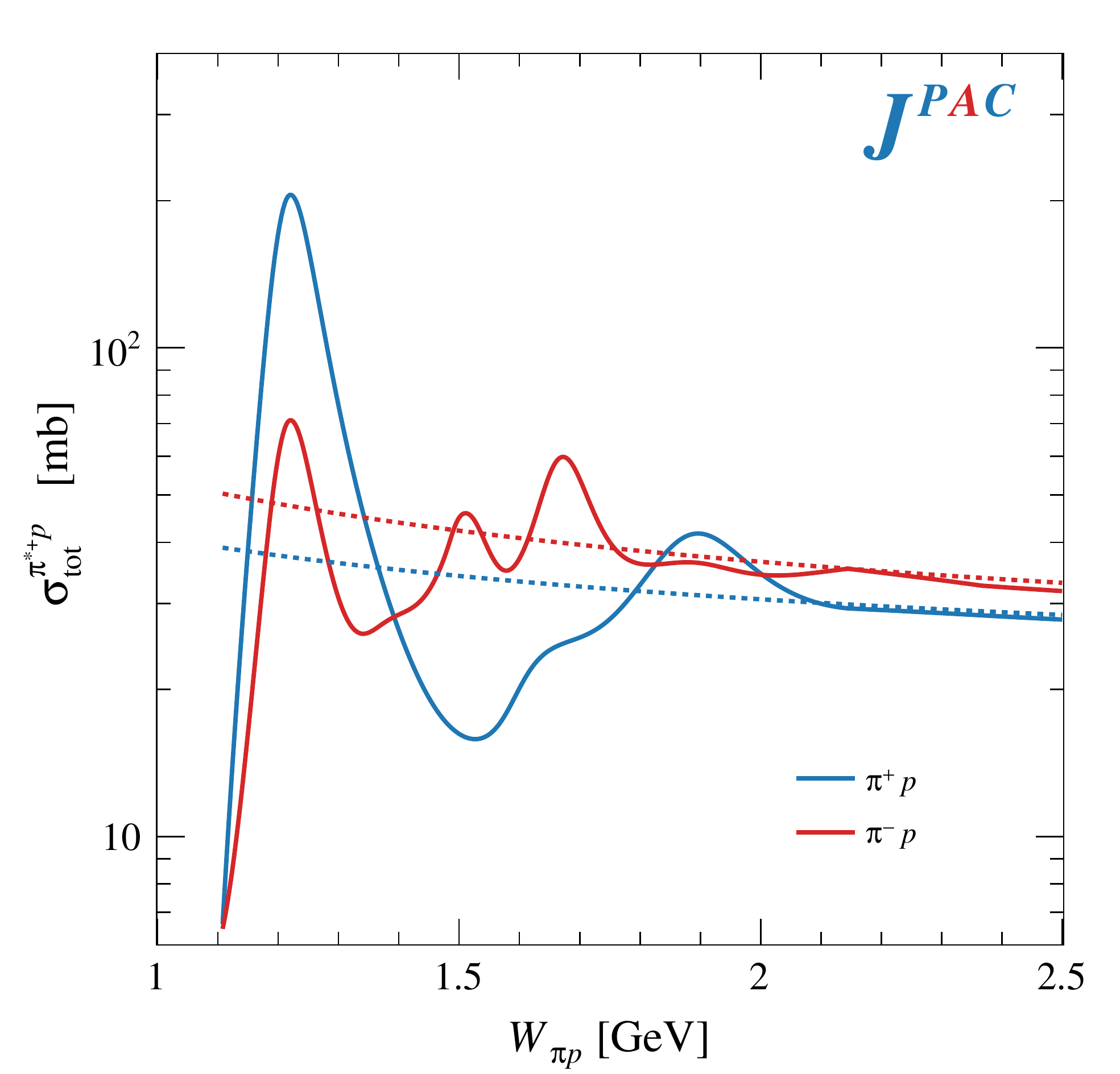}
        \caption{Low-energy behavior of the pion-nucleon elastic scattering as a function of the $\pi p$ invariant mass, $W_{\pi p} = \sqrt{s_{\pi p}}$. Solid lines are calculated with \cref{eq:sigmatot_main} with the amplitudes from~\cite{Mathieu:2015gxa}, while dashed are the high-energy parameterization \cref{eq:sigmatot_PDG}~\cite{ParticleDataGroup:2016lqr,COMPETE:2002jcr}.}
        \label{fig:sigmatot}
    \end{figure}
The SAID partial waves are provided up to $\M^2 \sim 4.5\gev$. At higher $\M^2$ 
 these are matched to a Regge-motivated parameterization which incorporates exchange physics relevant for the high energy regime. From \cref{eq:2-2_optical}, the cross section is calculated from $t$-channel isospin amplitudes:
    \begin{align} 
        \label{eq:sigmatot_main}
        \sigma_\text{tot}^{\pi^\pm p}(\M^2) = 
        \frac{\Im \left[C^{(+)}(\M^2, 0) \mp C^{(-)}(\M^2, 0)\right]}{p_\text{lab}(\M^2) \sqrt{s_0}}
       ~,
    \end{align}
where $C^{(\pm)}(s_{\pi p}, t_{\pi p})$ are the invariant, $t$-channel isoscalar and isovector $\pi p \to \pi p$ amplitudes,  with energy squared $s_{\pi p} = \M^2$, momentum transfer $t_{\pi p}$, and $p_\text{lab}(\M^2) = \lambda^{1/2}\!\left(\M^2, m_p^2, m_\pi^2\right)/2m_p$ the pion momentum in the proton rest frame. 
At larger $\M^2$, the amplitudes $C^{\pm}$ are taken to be the sum of Regge exchanges in the $t$-channel with the isoscalar corresponding to Pomeron and $f_2$ exchange while the isovector is dominated by $\rho$ exchange:
    \begin{align}
        \label{eq:Camps_Regge}
        C^{(+)}(\M^2, 0) &= A^\mathbb{P}(\M^2,0) + A^f(\M^2, 0)~, \nonumber \\
        C^{(-)}(\M^2,0) &= A^{\rho}(\M^2, 0)~.
    \end{align}
Here each scalar amplitude  $A^{k}(s_{\pi p}, t_{\pi p})$ takes a simple Regge pole form which in the forward direction is given by:
    \begin{align}
        \label{eq:regge_pole}
        A^{k}(&\M^2,  0) = - c^k_0 \; 
        \xi_k(0) \, \Gamma(n_k - \alpha_k(0)) \, \left(\frac{\hat{\nu}}{\sqrt{s_0}}\right)^{\alpha_k(0)} ~,
    \end{align}
with the (reduced) crossing-symmetric variable 
    \begin{equation}
        \label{nu}
        \hat{\nu} =  \nu(\M^2, 0) = \frac{\M^2 -  m_p^2 - m_\pi^2}{ 2 \, m_p}~,
    \end{equation}
which is the energy of the pion in the nucleon rest frame.
The $f_2$ trajectory is assumed to be degenerate with  the $\rho$ one with opposite signature. We take a simple linear form for all trajectories such that at $t_{\pi p} =0$ we only require the intercept $\alpha(0)$. 
The factor of $(n-\alpha(0))$ appearing inside the Gamma function is implemented with $n = (\tau + 1)/2$ to remove the ghost pole at $\alpha(t_{\pi p}) = 0$ for the $f_2$ and Pomeron exchanges in the more general expression. The parameters used are summarized in \cref{tab:sigmatotparams}.

\begin{table}
    \caption{Parameters for Reggeon contributions to $\sigma_\text{tot}^{\pi p}$ in  \cref{eq:regge_pole} and taken from~\cite{Mathieu:2015gxa}.}
        \label{tab:sigmatotparams}
        \begin{ruledtabular}
            \begin{tabular}{c c c c}
                  &  $\tau$ &   $\alpha(0)$  & $c_0$  \\
                \hline 
                $\mathbb{P}$ & $+$ & $1.075$  & $23.89$  \\
                $f$ & $+$ & \multirow{2}{*}{$0.49$}  &  $71.35$  \\
                $\rho$ & $-$ & & $5.01$
            \end{tabular}
        \end{ruledtabular}
\end{table}

The total cross section as a function of $\M^2$ is shown  in \cref{fig:sigmatot}. As a comparison we additionally plot the simpler phenomenological parameterization from  the PDG~\cite{COMPETE:2002jcr,ParticleDataGroup:2016lqr} the so-called $HPR_1R_2$ model, which takes the form:
    \begin{align}
        \label{eq:sigmatot_PDG}
        \sigma^{\pi^{\pm} p}_\text{tot}(\M^2) &= P + H  \, \log^2\left(\frac{\M^2}{M_\mu^2}\right)
         \\
        &\qquad + R_1 \, \left(\frac{\M^2}{M_\mu^2}\right)^{-\eta_1} \mp R_2\left(\frac{\M^2}{M_\mu^2}\right)^{-\eta_2}~.
        \nonumber
    \end{align}
Here $M_\mu^2 = (m_\pi + m_p + \mu)^2$ and the parameters $H = 0.272$ mb, $\mu = 2.1206\gev$, $\eta_1 =  0.4473$, and $\eta_2 =  0.5486$ are process independent. For $\pi p$ scattering the remaining parameters are $P = 18.75$, $R_1 = 9.56$, and $R_2 = 1.767$, all in units of mb. We consider this model to examine the impact of the low-energy resonance region in the production rates.

In order to incorporate the off-shell pion in a minimal way we will assume that the dependence on the virtuality enters only through a kinematic change of phase space factors. For the low-energy regime we modify \cref{eq:sigmaL_pip} with a factorized form:
    \begin{equation}
        \label{eq:sigma_rescale}
        \sigma^{\pi^{*\pm}p}_\text{tot}(t, \M^2) = \sum_L \, R^{\pi^*}_L(t, \M^2) \, \sigma_L^{\pi^\pm p}(\M^2)
       ~,
    \end{equation}
where
    \begin{equation}
        R_L^{\pi^*}(t, \M^2) = \left(\frac{p_{\pi^*}}{p_\pi}\right)^{2L} = \left(\frac{\lambda(\M^2, t, m_p^2)}{\lambda(\M^2, m_\pi^2, m_p^2)}\right)^{L} 
    \end{equation}
is the ratio of barrier factors of individual partial waves. We note that in the limit $\M^2 \gg |t|$ the rescaling ratio tends to unity and there is no need to apply it to the high-energy, Regge part in \cref{eq:Camps_Regge}. 

The rescaling in \cref{eq:sigma_rescale} numerically involves large cancellations when evaluated very close to threshold, in particular for high-$L$ waves. In the numerical studies below, we replace the value $L\to  \min (L, L_\text{max})$ in the rescaling factor $R_L^{\pi^*}$, with the value of $L_\text{max}$ chosen to provide the appropriate rescaling to the dominant waves while keeping the higher waves numerically stable near threshold. Because  the  resonance peaks appear mainly in the $S$- and $P$-waves, with higher waves contributing primarily to the intermediate $\M^2$ region near the matching point, we find that $L_\text{max} = 3$ is sufficient for all processes considered. 
\subsection{Relation to triple Regge amplitude}
\label{sec:tripleregge}
\begin{figure*}
        \centering
        \includegraphics[width=.8\textwidth]{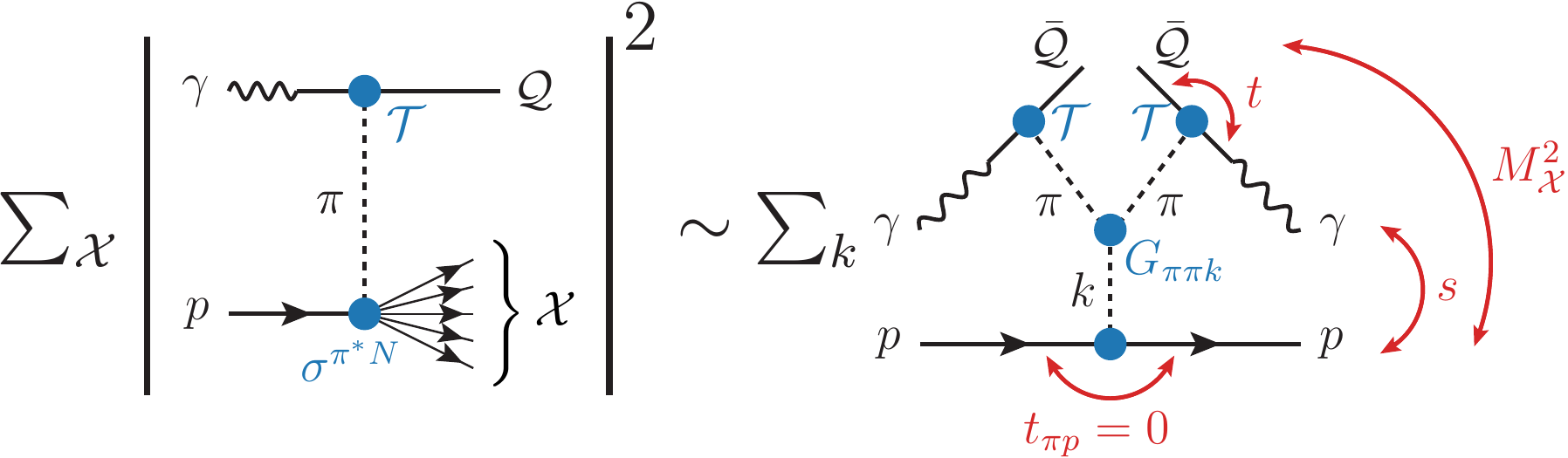}
        \caption{Diagrammatic representation of the triple Regge formula of \cref{eq:triple_regge}. }
        \label{fig:triple}
    \end{figure*}

The model in \cref{eq:dtsigma_gen} has been considered in the past  in the context of the triple Regge  formula  (see for example~\cite{Field:1974fg,Ganguli:1980ec}), which takes the form:
    \begin{align}
        \nonumber
        \dtsig &\left(\gamma p \to \mQ^\pm \mN\right)=
         \\
        \sum_{k} \;\; & \frac{G_{\pi\pi k}^\mp(t)}{\pi \,s_0 \, s}  \left(\frac{s}{\M^2}\right)^{2\alpha_\pi(t)} \, \left(\frac{\M^2}{s_0}\right)^{\alpha_k(0)}
       ~,        \label{eq:triple_regge}
    \end{align}
schematically represented in \cref{fig:triple}. Here $k$ labels the possible Regge exchanges in the bottom vertex, and $G_{\pi\pi k}$ is a triple Regge coupling, often parameterized with a phenomenological exponential form. The scale $s_0$ is customarily taken to $1\gevsq$. This form is expected to be reliable in the triple Regge kinematic region $s \gg \M \gg |t|$,  where the particle $\mQ$ carries large momentum in the near-forward direction, \ie when the fraction of longitudinal momentum $x \sim 1$. 
Since most of the literature focuses on this limit, it is worth considering how \cref{eq:dtsigma_gen} compares to the phenomenological formula in \cref{eq:triple_regge}. 

If we consider  \cref{eq:dtsigma_gen} in the triple Regge  kinematics,  we first note that the flux ratio $K \to (\M^2 / s)$ and we may use the Reggeized form in \cref{eq:mP_R} for the pion propagator. We must then consider $\sigma^{\pi^* p}_\text{tot}(t, \M^2)$ in this kinematics. From \cref{eq:regge_pole}, as $\M^2\to \infty$ we have $R_{L}^{\pi^*} \to 1$ and the Reggeon pole form \cref{eq:regge_pole} reduces to:
    \begin{equation}
        \label{eq:Aamps_highM2}
        \Im A^k(\M^2, 0) \to \gamma_k \left(\frac{\M^2}{s_0}\right)^{\alpha_k(0)},
    \end{equation}
where $k=\mathbb{P},f_2,\rho$ and the
coupling constants are given by
    \begin{equation}
        \gamma_k = \frac{\pi}{2}  \, \frac{ 
        c^k_0 }{\Gamma\left(\alpha_k(0) - n_k +1\right) } \left(\frac{\sqrt{s_0}}{2\, m_p}\right)^{\alpha_k(0)}~.
    \end{equation}
Then, with \cref{eq:Aamps_highM2}, we may write \cref{eq:dtsigma_gen} as:
    \begin{equation}
        \sigma^{\pi^{*\pm} N}_\text{tot}(\M^2) \to \sum_{k} \, \frac{2 m_p}{\sqrt{s_0}} \frac{\gamma_k f^\pm_k}{s_0} \left(\frac{\M^2}{s_0}\right)^{\alpha_k(0) -1 }~.
    \end{equation}
where $f^\pm_\rho = \mp 1$ and $f^\pm_{f_2,\mathbb{P}} = +1$ depending  on the pion charge. 
Putting it  all together, we can see  that \cref{eq:sigmatot_main} in the triple Regge limit reduces  precisely the form of \cref{eq:triple_regge} with the triple Regge vertex function given by
    \begin{align}
        G_{\pi\pi k}^\pm&(t) = f^\pm_k \; \frac{ m_p \sqrt{s_0} \, \gamma_k}{8\pi^2} 
        \nonumber \\
        &\qquad\times\Big|\alpha'\, T_\pi(t) \,\xi_\pi(t) \, \Gamma(-\alpha_\pi(t))\Big|^2
        ~.
    \end{align}

Seeing the emergence of triple Regge behavior in the appropriate limit is reassuring, as the formula \cref{eq:dtsigma_gen} generalizes the semi-inclusive cross section by loosening the requirement of large $\M^2$ and allows us to consider the fixed spin analog relevant for near threshold production, \ie with small $s$ and $\M^2$. 

\subsection{Exclusive $\mQ^- \Delta^{++}$ and  $\mQ^+ \Delta^{0}$ 
 production}
\label{sec:deltapp}

Another consistency test for the semi-inclusive cross section formula in \cref{eq:dtsigma_gen} is the opposite limit, $\M \sim M_\text{min}$. We may consider the production of a  meson $\mQ$ together with a $\Delta$ baryon, which  dominates the low-energy $\pi N$ spectrum. 
This is particularly true for the $\pi^+ p$ scattering, where the $\Delta^{++}$ is the only resonance in the mass region $\M \lesssim 1.5\gev$. 

Additionally, in this $\pi N$ isospin-channel there is no analogous exclusive reaction, meaning the $\gamma p \to \mQ^{-} \, \Delta^{++}$ reaction is already contained within the cross section, \cref{eq:dtsigma_gen}.
Calculating the exclusive reaction with effective Lagrangian methods as in Ref.~\cite{Albaladejo:2020tzt} should lie strictly below the prediction for the 
 inclusive cross section and saturate the low energy regime. Although we will focus specifically on the $\mQ^-$ case, the discussion in this section is directly  applicable for the opposite charge with a relative factor associated with isospin-projection:
    \begin{equation}
        \frac{\sigma(\gamma p \to \mQ^+\Delta^0)}{\sigma(\gamma p \to \mQ^-\Delta^{++})} = \frac{1}{3}~.
    \end{equation}

Because $P$-wave $\pi^+p$ scattering is  dominated by the $\Delta^{++}$, a straightforward way to consider $\mQ^{-}\Delta^{++}$ production is to restrict the total cross section in \cref{eq:dtsigma_gen} to only the $P$-wave component by replacing 
    \begin{equation}
        \label{eq:sigma_fixL}
        \sigma_\text{tot}^{\pi^{\ast+}p}(t, \M^2) \to R_{L=1}^{\pi^\ast}(t, \M^2) \; \sigma_{L=1}^{\pi^+ p}(\M^2)~.
    \end{equation}

Alternatively, in analogy to the exclusive formalism of Ref.~\cite{Albaladejo:2020tzt}, we can include the spin-3/2 baryon by changing the bottom, $\pi\Delta N$ vertex for which we take a simple effective Lagrangian as in~\cite{Nam:2011np}:
    \begin{equation}
        \mathcal{L}_{\pi N \Delta} = \frac{g_{\pi N\Delta}}{m_\pi} \, \bar{\Delta}^\mu \, \partial_\mu \pi \, N + \text{H.c.}~,
    \end{equation}
or equivalently the bottom vertex:
    \begin{equation}
        \mathcal{B}_{\lambda_N,\lambda_\Delta} = \frac{ i \,g_{\pi N\Delta}}{m_\pi} \, \bar{u}^\mu(p^\prime, \lambda_\Delta) \, k_\mu \, u(p, \lambda_N)~,
    \end{equation}
where $k$ is the pion momentum, $k = p - p^\prime$ and $u^\mu$ is the 
Rarita-Schwinger spinor:
    \begin{align}
        u^\mu(p, \lambda) &=  
        \nonumber \\ 
        \sum_{m_1, m_2} &\braket{1, m_1; \tfrac{1}{2}, m_2}{\tfrac{3}{2},\lambda} \, \epsilon^\mu(p, m_1) \, u(p, m_2)~.
    \end{align}
The coupling $g_{\pi N \Delta}$ is calculated assuming the width to be saturated by the $\pi N$ final state. With  $\Gamma_{\Delta} = 120\mev$ and $m_\Delta = 1.23\gev$, this leads to $g_{\pi N \Delta} = 2.10$. 

We may thus calculate the $\gamma p \to \mQ\Delta$ amplitude assuming a stable $\Delta$ in the final state. To directly compare to the cross section in \cref{eq:dtsigma_gen} we incorporate the $\Delta \to \pi p$ lineshape via
    \begin{align}
        \label{eq:delta_bw}
        \sigma(\gamma p &\to \mQ^- \, \pi^+ p) = 
        \\
        &\int_{M^2_\text{min}}^\infty dM^2 \, \sigma(\gamma p \to \mQ^- \, \Delta^{++}) \, d_{\Delta\to\pi p}(M^2)~,
        \nonumber
    \end{align}
where the $\mQ\Delta$ cross section is calculated at fixed $s$ as a function of $M^2$. Because of the proximity of the $\pi p$ threshold we use a simple Breit-Wigner-like distribution proposed in Ref.~\cite{Giacosa:2021mbz} which is shown to provide a good  description of the $\pi p$ mass distribution in the $\Delta$ mass region:
    \begin{align}
        d_{\Delta\to\pi p}(M^2) = \frac{1}{\pi} \frac{\rho(M^2) \, \tilde{\Gamma}_\Delta}{[M^2 - m_\Delta^2]^2 + [\rho(M^2) \, \tilde{\Gamma}_\Delta]^2}~,
    \end{align}
with $\rho(M^2) = \sqrt{M^2 - M^2_\text{min}}$ and $\tilde{\Gamma}_\Delta = \Gamma_\Delta \, m_\Delta / \rho(m_\Delta^2)$. Interestingly, this function is normalized across the mass distribution, obeying
\begin{equation}
        \int_{M^2_\text{min}}^\infty dM^2  \, d_{\Delta\to\pi p}(M^2) =1~.
\end{equation}
Both \cref{eq:delta_bw,eq:sigma_fixL} are expected to yield similar results with minor differences related to the lineshape assumed for the $\Delta$. 

\section{Numerical results}
\label{sec:numerics}
    \begin{figure}
        \centering
        \includegraphics[width=\columnwidth]{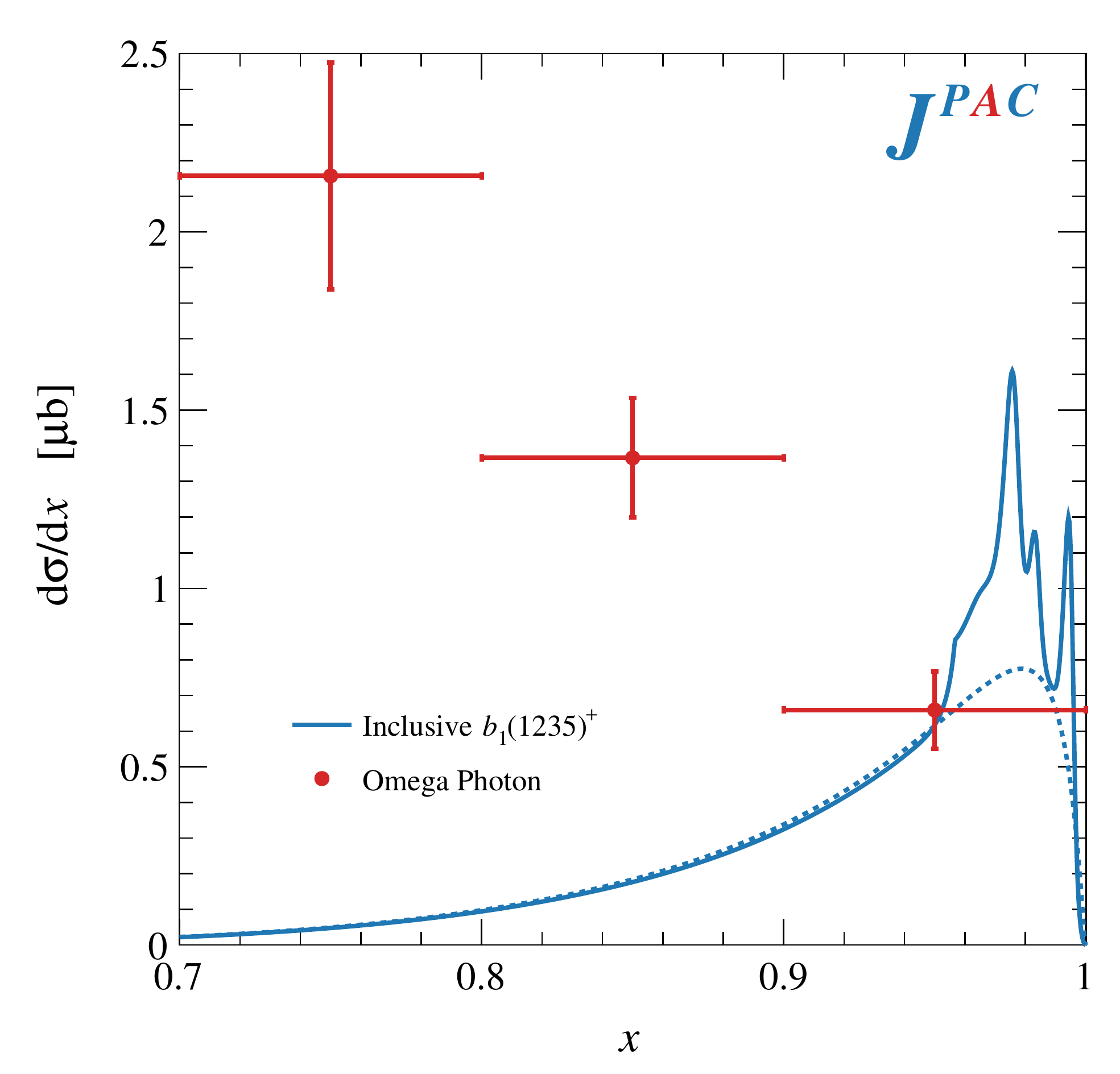}
        \caption{Inclusive $b_1(1235)^+$ production at $W_{\gamma p} = 8.7\gev$ using the Reggeized pion exchange. The solid curves are the inclusive cross section using the parameterization in \cref{eq:sigma_rescale} including nucleon resonances, while the dashed curves calculate the same process using the Regge-only parameterization of \cref{eq:sigmatot_PDG}. When averaged over the highest bin, both curves are consistent with the experimental point within errors. Data  from~\cite{OmegaPhoton:1984xqe}.}
        \label{fig:b1_dsigmadx}
    \end{figure}
Experimental data on the photoproduction of exotic quarkoniumlike states is virtually nonexistent. In order to benchmark the predictions for $Z$ states, we first consider the axial-vector analog in the light sector, the charged $b_1(1235)$ which has been looked at in the kinematic region of interest by the OmegaPhoton collaboration~\cite{OmegaPhoton:1984xqe}. The applicability of the same amplitudes for both light and heavy meson production constitutes the primary model assumption, \ie through the use of VMD which has recently been criticized~\cite{Xu:2021mju}. As noted in Ref.~\cite{JPAC:2021rxu}, however, VMD predicts roughly the correct size of the ratio $\Gamma(\chi_{c2} \to \gamma \,J/\psi) / \Gamma(\chi_{c2} \to \gamma \gamma)$, so it seems an appropriate method to obtain at least order-of-magnitude estimates. In any case, we note this assumption affects only the top vertex which is the same as in the exclusive analysis. Thus considering the $b_1$ still serves as a test of the inclusive extension of the pion exchange process which in principle is independent of the microscopic nature of $\mQ$. 

We begin with the comparison with the differential cross section data in Ref.~\cite{OmegaPhoton:1984xqe}. The energy range covered by the measurement is $E_\gamma = 25$--$55\gev$ and for simplicity we take the midpoint center-of-mass energy, $W_{\gamma p} \sim 8.7\gev$. At these energies with respect to the $b_1 N$ threshold, the triple Regge behavior is expected to be dominant and we use exclusively the Reggeized form of the pion propagator, \cref{eq:mP_R}, with the high energy approximated kinematics as explained in \cref{app:kinematics}.

The only undetermined parameter is the $g_{\gamma b_1 \pi}$ coupling governing the strength of the top vertex. We use the effective Lagrangian formalism previously considered in Ref.~\cite{Albaladejo:2020tzt} for the $\gamma \pi b_1$ vertex to extract this coupling from its decay widths. Luckily the radiative decay width for the $b_1$ is known and we may extract the coupling without relying on VMD. For $\Gamma(b_1 \to \pi \gamma) = 230\kev$~\cite{Collick:1984dkp} we have $g_{\gamma b_1 \pi} = 0.24$.

For comparison purposes, we calculate the inclusive cross section with the pion-nucleon interaction described with the full $\pi^* N$ cross section \cref{eq:sigmatot_main}, as well as with the Regge-only parameterization of \cref{eq:sigmatot_PDG}. These are shown in \cref{fig:b1_dsigmadx} compared to the data points in the highest $x$ bins. We see a good agreement of both the models in the highest bin. To make the comparison more quantitative we may calculate the average cross section from each curve in this bin, yielding 0.69 and 0.56 $\mu$b for the model including nucleon resonances and the Regge-only parameterization respectively, both consistent with OmegaPhoton within uncertainties. 
The agreement between the two values is expected, because at this relatively large center-of-mass energy the nucleon resonances are squeezed in a small portion of the phase space. 

Away from the highest bin we note that single pion exchange 
 severely underestimates the production. Indeed, since at high energies the upper limit of integration $t$ is proportional to $1-x$ and the pion exchange is exponentially suppressed with $|t|$, the integral is also exponentially suppressed in $1-x$. This suggests that the triple Regge contribution becomes quickly irrelevant already at $x<0.9$, although other top exchanges than the pion should be added.  Summarizing, we find good agreement with data in the region of validity of the model, and a trend toward zero away from this region. We may thus consider the cross section calculated in this formalism as a conservative lower bound of the total inclusive production rate.

\begin{figure}
        \centering
        \includegraphics[width=\columnwidth]{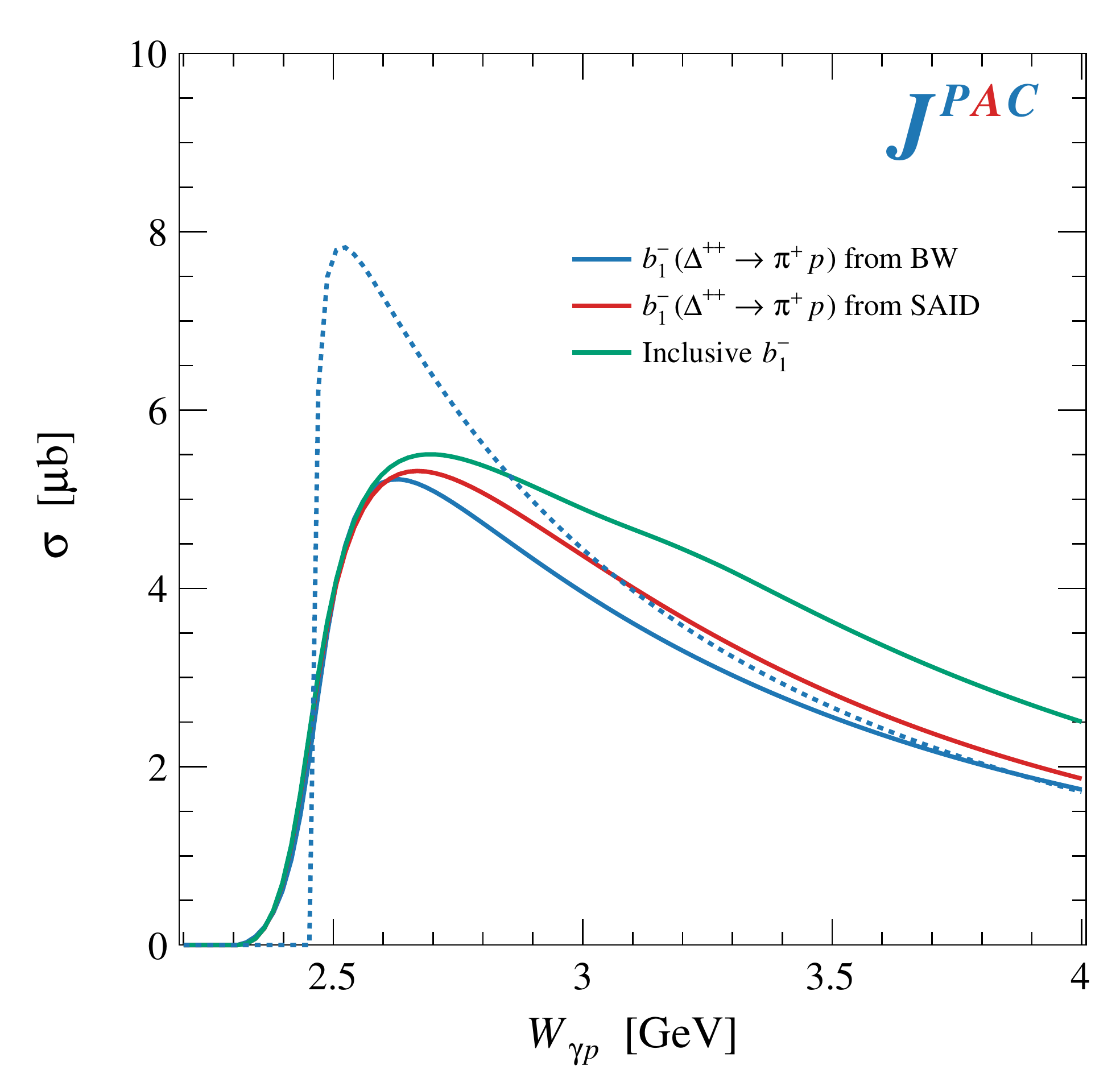}
        \caption{Inclusive $b_1(1235)^-$ production as compared to exclusive $b_1^- p\pi^+$ production through an intermediate $\Delta^{++}$ as a function of center-of-mass energy. The $\Delta\to\pi p$ decay is incorporated via a BW shape using \cref{eq:delta_bw} or from restricting the SAID PW sum in \cref{eq:sigma_fixL}. The dashed line is the exclusive reaction calculated with the effective Lagrangian assuming a stable $\Delta$ baryon.} 
        \label{fig:b1_deltapp}
    \end{figure}

\begin{figure}
        \centering
        \includegraphics[width=\columnwidth]{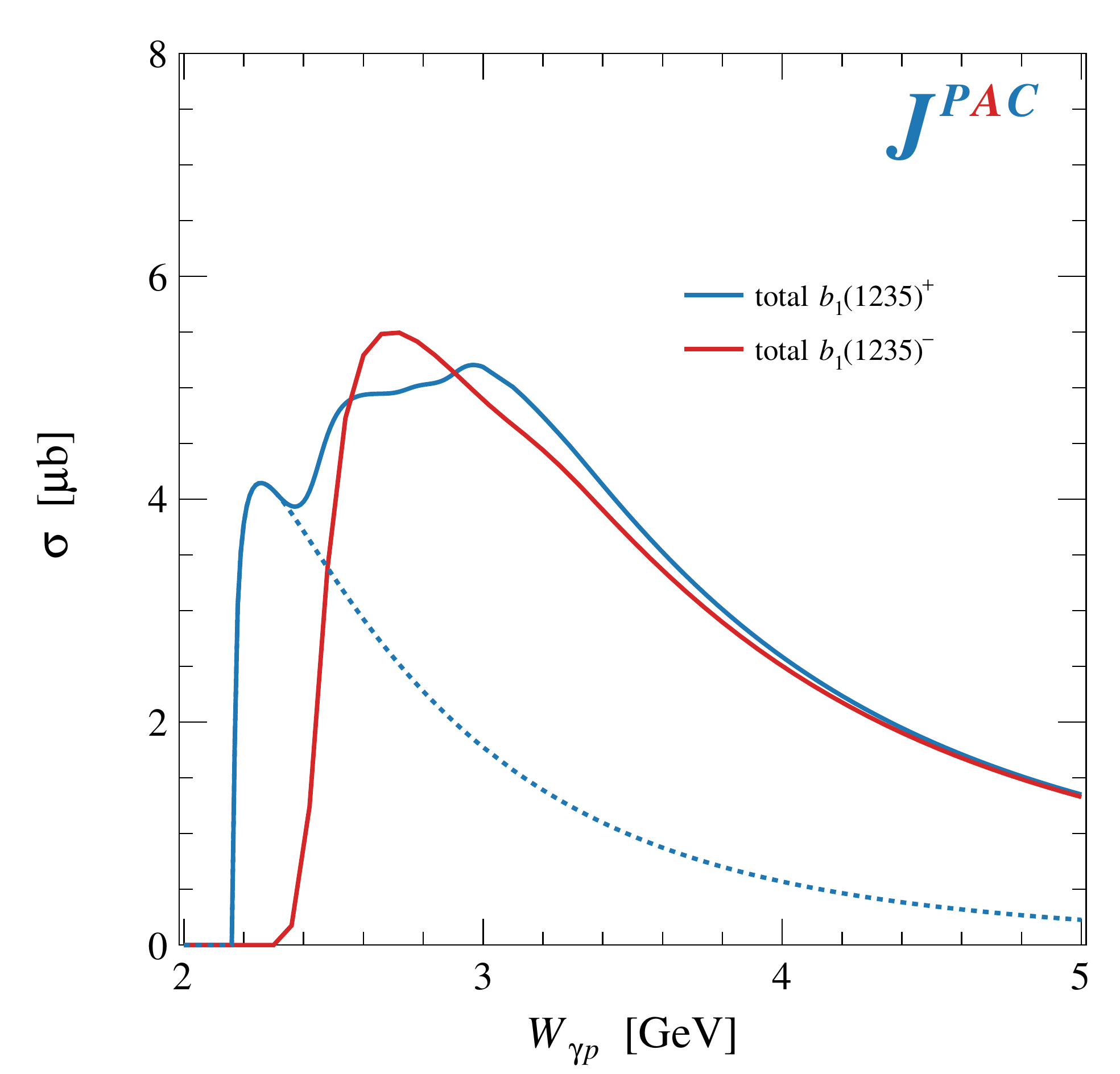}
        \caption{Total $b_1(1235)$ production as a function of $W_{\gamma p}$ with a fixed-spin pion exchange. The $b_1^+$ curve includes the sum of the inclusive contribution \cref{eq:dtsigma_gen} and the exclusive process (dashed) as calculated in~\cite{Albaladejo:2020tzt} which lies below the $\pi N$ threshold. The $b_1^-$ production does not have a corresponding exclusive analog. }
        \label{fig:b1_totals}
    \end{figure}

By integrating over $x$,  
we may also investigate the near-threshold behavior as a function of invariant mass. For energies $W_{\gamma p} \lesssim 5\gev$ we use instead the fixed-spin pion exchange model of \cref{eq:mP_FS}. We compare the inclusive prediction for $b_1(1235)^-$ production  to the explicit $b_1^- \Delta^{++}$ using the formalism in \cref{sec:deltapp}. The comparisons between the exclusive $\Delta^{++}$ prediction with and without inclusion of the subsequent $\Delta\to\pi N$ decay are shown  in \cref{fig:b1_deltapp} compared to the inclusive production. 
We see that the  unstable $\Delta$ curve saturates the inclusive production up to the nominal $b_1 \Delta$ threshold, after which the contribution of other resonances 
 become relevant. We also see good agreement between the two methods of incorporating the $\Delta^{++}$ decay. To this end in all subsequent numerical studies we 
   consider the $\Delta\to\pi p$ decay by restricting the total cross section in \cref{eq:sigma_fixL} to the SAID $I=3/2$, $L=1$ partial wave, since it incorporates more accurately the $\Delta^{++}$ lineshape. The inclusive cross sections of both charged $b_1^\pm$ are shown in \cref{fig:b1_totals}.

    \begin{figure*}[t]
        \centering      
        \includegraphics[width=\columnwidth]{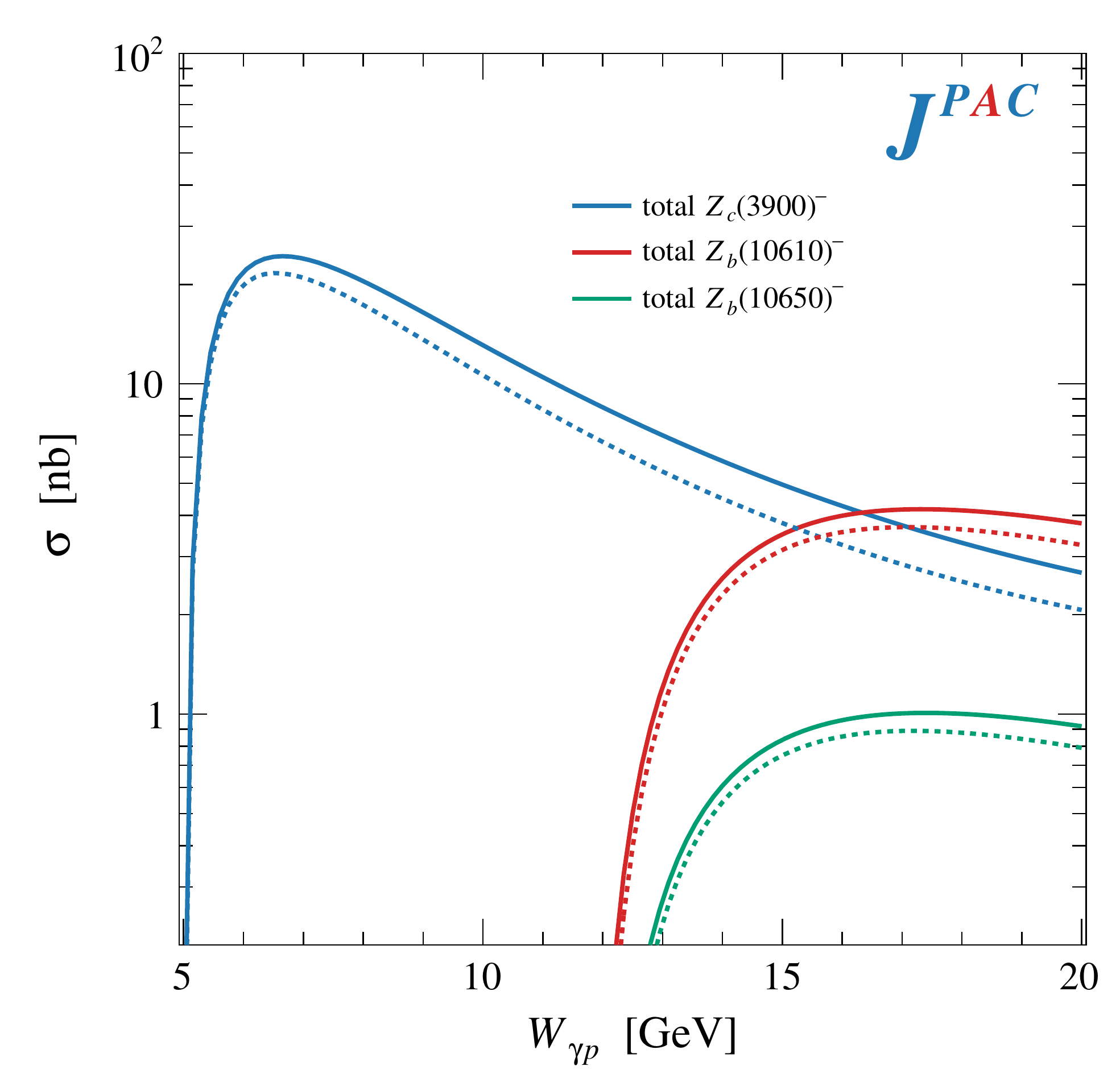}
        \includegraphics[width=\columnwidth]{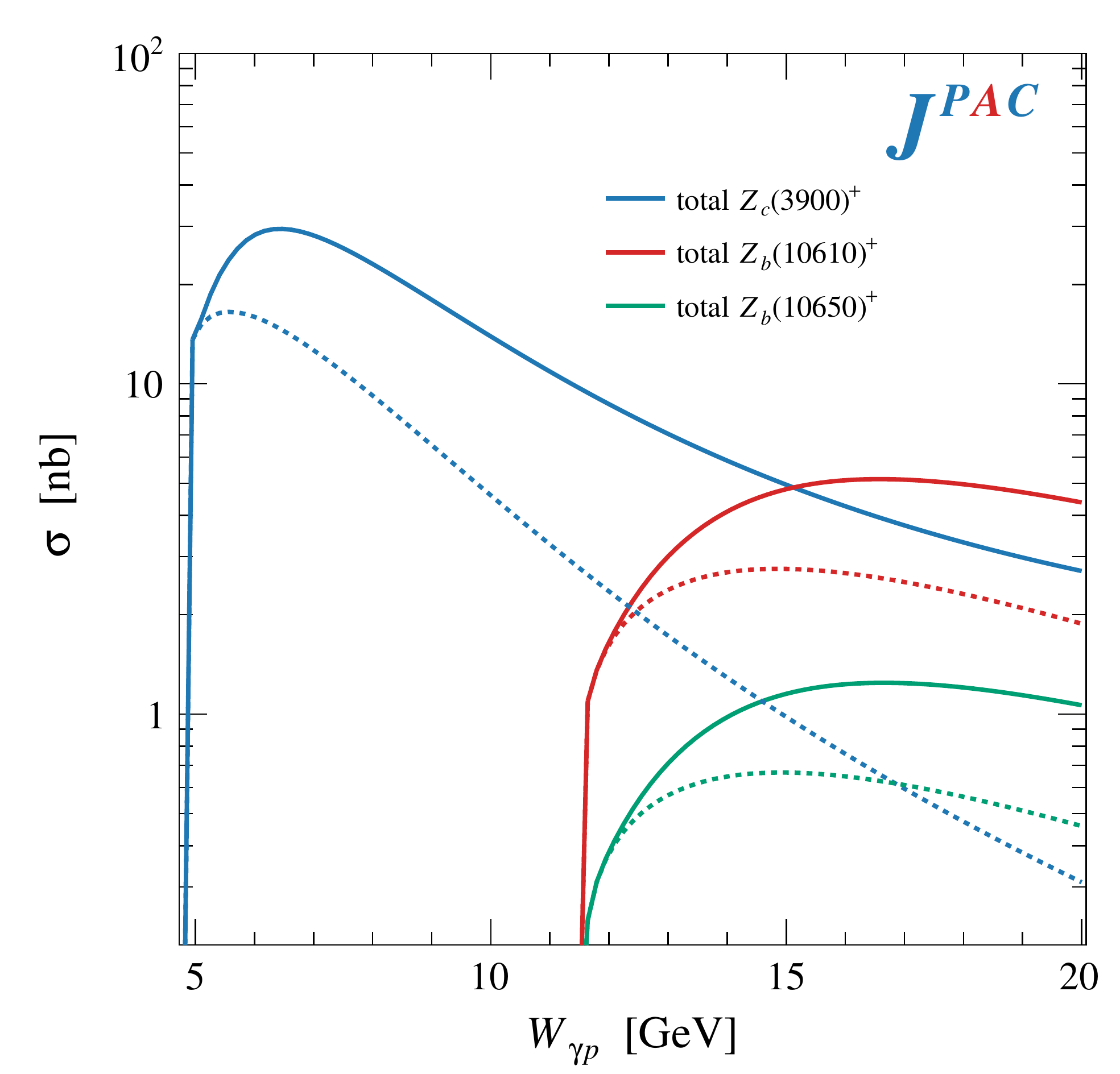}
        \caption{Total cross section predictions for charged, charmoniumlike $Z$-states near-threshold via fixed-spin pion exchange. Left panel: Total inclusive $Z^-$ cross sections (solid lines) as compared with the exclusive $\gamma p \to Z^- \Delta^{++} \to Z^{-} \pi^+ p$ cross section (dashed lines) as described in \cref{sec:deltapp}. Right panel: Total $Z^+$ cross sections (solid lines), which include the sum of the inclusive cross section and the exclusive nucleon pole contribution. This latter contribution is calculated as in Ref.~\cite{Albaladejo:2020tzt} and is shown explicitly in dashed lines.}
        \label{fig:Z_totals}
    \end{figure*}
    \begin{table*}[t]
        \centering
        \caption{High-energy production cross sections of the total inclusive process $\gamma p \to \mQ\mN$ as a function of $W_{\gamma p}$, compared to the exclusive process. The exclusive cross sections fall asymptotically to zero while the inclusive approaches a constant as described in the text.}
        \label{tab:asymptotics}
        \begin{ruledtabular}
        \begin{tabular}{c  c c c | c c c}
            & \multicolumn{3}{c}{$\sigma(\gamma p \to \mQ^{\pm} \, \mN)$ \, [pb]}  & \multicolumn{3}{c}{$\sigma(\gamma p \to \mQ^+ n)$ \, [pb]} \\
            \hline
            $\mQ$ & $30\gev$ & $60\gev$ & $90\gev$ & $30\gev$ & $60\gev$ & $90\gev$ \\
            \hline
            $b_1(1235)$       & $60\cdot10^{3}$ & $60\cdot10^{3}$ & $61\cdot10^{3}$  & 43 & 2.3 & $<10^{-8}$ \\
            $Z_c(3900)$  & 187 & 146 & 140 & 19 & 1.0 & $<10^{-8}$ \\
            $Z_b(10610)$  & 163 & 15 & 5 & 150 & 10 & $<10^{-8}$ \\
            $Z_b(10650)$  & 40 & 4 & 1 & 37 & 2.4 & $<10^{-8}$ \\ 
        \end{tabular}
        \end{ruledtabular}
    \end{table*}

\begin{figure}
        \centering
        \includegraphics[width=\columnwidth]{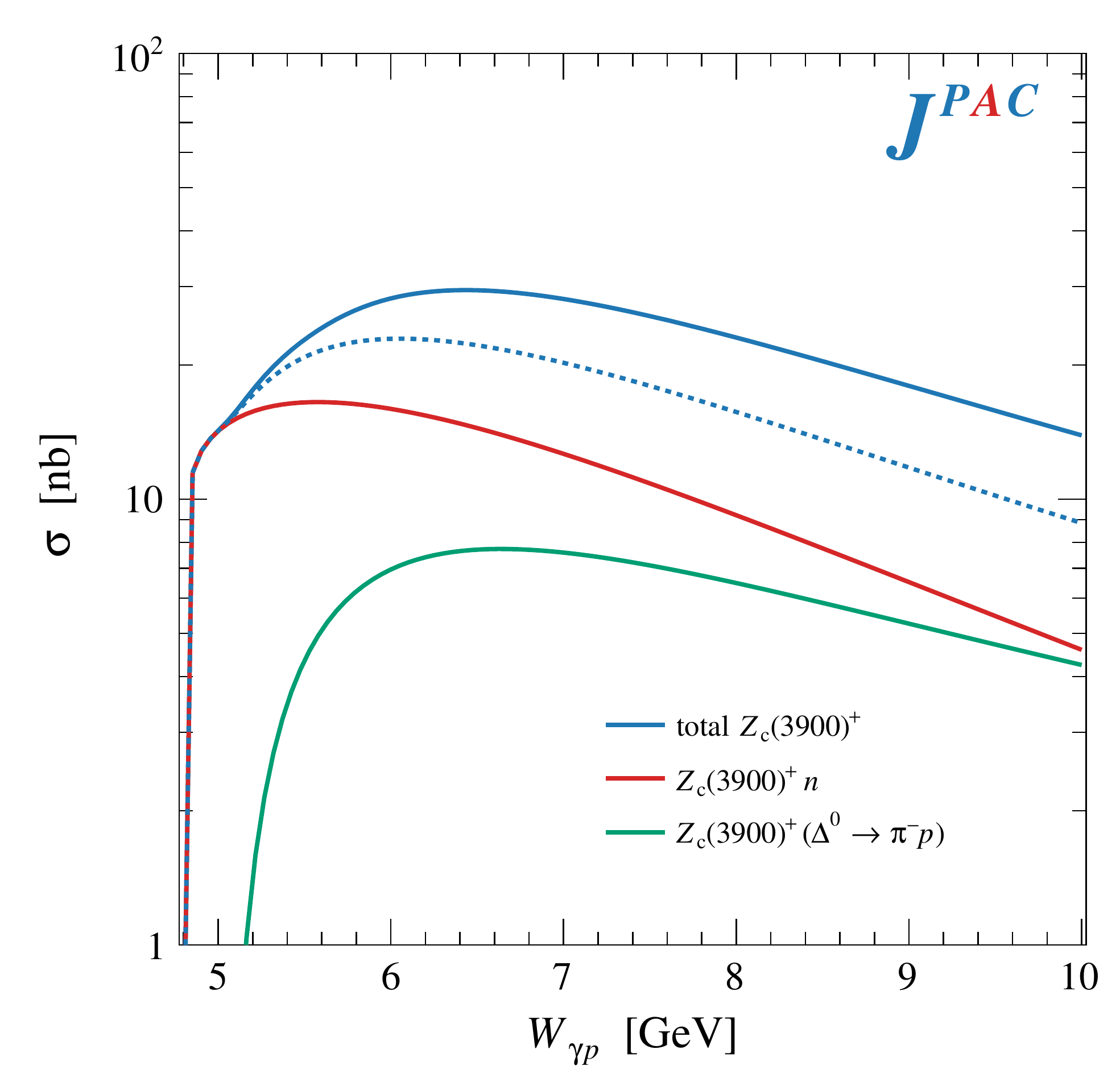}
        \caption{Contributions to the total cross section of near-threshold $Z_{c}(3900)^+$ production. The total curve represents the sum of the full inclusive contribution and the exclusive reaction. The dashed line corresponds to the sum of only $Z_c^+ n$ and $Z_c^+ \Delta^0$ contributions.}       \label{fig:Zcm_compare}
    \end{figure}

We now turn to the quarkoniumlike states in the hidden charm and bottom sectors. We use the couplings that were previously calculated from the observed hadronic decays of the $Z$-states within the VMD model,  $g_{\gamma Z \pi} \times 10^{2} = 5.17, 5.8,$ and $2.9$ for $Z_c, Z_b$ and $Z_b^\prime$ respectively~\cite{Albaladejo:2020tzt}. The total near-threshold production cross sections for the two charged  $Z^\pm$ states are shown in \cref{fig:Z_totals}. Examining the $Z^-$ inclusive cross sections, we again note that it is dominated by the $\Delta^{++}$ resonance. On the other hand, the contributions from inelastic channels to $Z^+$ production is roughly the same size as the exclusive $Z^+n$ reaction enhancing the total production rate by a factor of 2 close to threshold. \Cref{fig:Zcm_compare} provides a more detailed picture of the contributing processes. We see that, unlike the $Z^-$ case, the $\Delta^0$ does not dominate the inclusive cross-section with contributions from other resonances being equally important. 
Further we see the asymptotic behavior of the inclusive cross section falls much slower than the $Z^+ n$ final state 
as the center-of-mass energy grows. In fact, from the comparison with the asymptotic triple Regge formula in \cref{sec:tripleregge}, we expect the curves to flatten out and grow slowly with energy, while the exclusive cross section decreases~\cite{Albaladejo:2020tzt}. Explicit comparison of inclusive production compared to the respective exclusive reaction at large energies is shown in \cref{tab:asymptotics}.  

\begin{figure}
    \centering
    \includegraphics[width=\columnwidth]{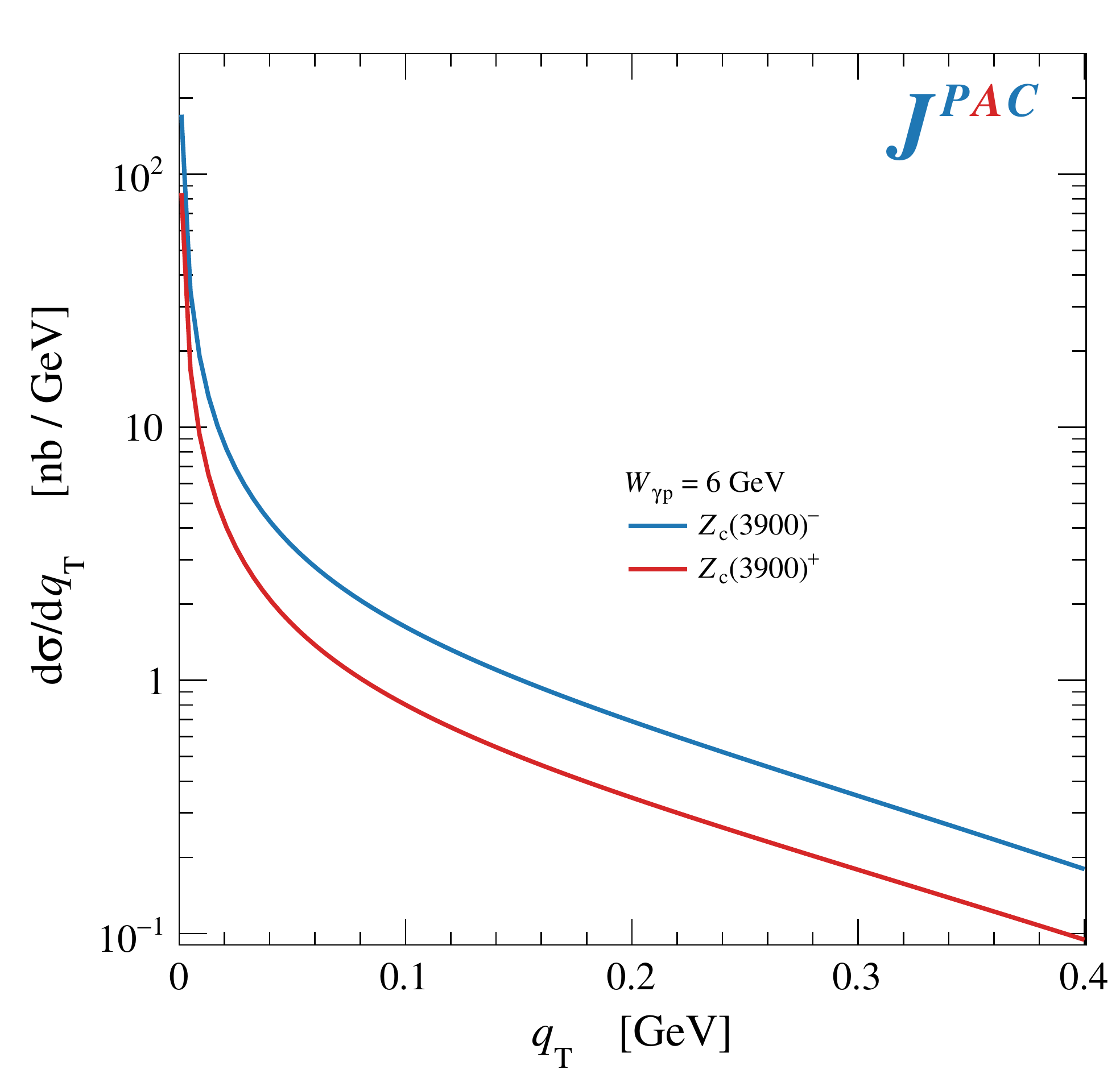}
    \caption{Differential distributions with respect to the transverse momentum of $Z_c(3900)$, i.e. $q_{T} = q_\text{max} \, y = q_f \, \cos\theta$, for near-threshold production.}
    \label{fig:Z_pT}
\end{figure}

In \cref{fig:Z_pT} we show the transverse momentum distribution of the cross section. The diffractive production mechanism contributes primarily to the small-$q_T$ region, meaning it may be expected to be predominant at near-threshold energies where the possible produced inclusive final states have small invariant mass.

\section{Conclusions}
\label{sec:conclusions}

We have calculated the semi-inclusive photoproduction rates of the axial-vector, charmonium-like $Z^\pm$ states. We focus on the diffractive kinematic region where charged pion exchange is assumed to be the dominant production mechanism. We use a formalism akin to the triple-Regge interaction model while incorporating the effects of nucleon resonances. By focusing on pion exchange we performed a series of benchmarks testing the predictions against data available for the analogous $b_1$ meson, as well as expectations from effective Lagrangian methods for small missing mass. 

The analysis indicates that the inclusion of semi-inclusive final states produces cross sections upwards of tens of nanobarn for the $Z_c$ and few nanobarn for $Z_b$ states. This comes from a  rough factor of 2 increase compared to the exclusive production in the near-threshold production. In particular the $\Delta$ resonance is found to play a large contributing role in the near-threshold production. It may thus be viable to search for these states in the \textit{exclusive} $Z^{+} \, \Delta^{0}$ and $Z^{-}\,\Delta^{++}$ mode which are of comparable size to the $Z^+\,n$ reaction. This feasibility will depend on the experimental setup's ability to reconstruct the $\Delta$ decay pions to which the formalism presented here may be a valuable tool for simulation.

Similar semi-inclusive electroproduction at electron-hadron facilities has been considered recently in~\cite{Yang:2021jof}. Herein production rates are calculated within the hadronic molecule interpretation of the $Z_c$ states as an $S$-wave $D\bar{D}^*$ bound state and therefore arise from the semi-inclusive production of constituents which are then rescattered. The initial constituent distribution is estimated with general-purpose MonteCarlo generators (Pythia). With no specific tuning, the charm production is predominantly due to hard scattering. Extrapolating this to the region close to threshold might severely underestimate the production rates.
COMPASS has measured upper limits for the $Z_c(3900)$ exclusive photoproduction cross sections at an average energy of $\left\langle W_{\gamma p}\right\rangle = 13.8\gev$ of $\sim 0.5$ nb, once branching ratios are taken into account~\cite{COMPASS:2014mhq}. Were this to be confirmed, this would imply an overestimate of a factor of $\sim 4$ with respect to our exclusive predictions, which could be due to the breaking of the VMD assumption, or to a dramatic dependence of the top coupling on the photon virtuality. An independent confirmation of such result is thus needed. 

The predictions presented here, while likely a lower-bound of the total expected semi-inclusive production of $\Zs$, do not assume any microscopic nature of the produced state. 
Further, the formalism presented here is naturally extendable to the semi-inclusive production of other exotic candidates, in particular the $X(3872)$, $Y(4260)$ and hidden-charm pentaquark states. These are assumed to be produced via exchanges with spin and thus the work presented here is a stepping stone to considering more complicated reactions relevant for the spectroscopy programs at future facilities.

\acknowledgments
This work was supported by the U.S.~Department of Energy under Grants
No.~DE-AC05-06OR23177 
and No.~DE-FG02-87ER40365, 
the U.S.~National Science Foundation under Grant 
No.~PHY-1415459. 
It was also supported by Deutsche Forschungsgemeinschaft (DFG) through the Research Unit FOR
2926 (project number 40824754). 
DW is supported by National Natural Science Foundation of China Grant No.~12035007
and the NSFC and the Deutsche Forschungsgemeinschaft (DFG, German Research Foundation) through the funds provided to the Sino-German Collaborative Research Center TRR110 ``Symmetries and the Emergence of Structure in QCD" (NSFC Grant No.~12070131001, DFG Project-ID~196253076-TRR~110). 
%
VM is a Serra H\'unter fellow and acknowledges support from the Spanish national Grant No. PID2019–106080 GB-C21 and PID2020-118758GB-I00.  
MA is supported by Generalitat Valenciana under Grant No. CIDEGENT/2020/002. 

\appendix
\section{Kinematics of semi-inclusive photoproduction}
\label{app:kinematics}
\begin{figure}[t]
        \centering
        \includegraphics[width=0.96\columnwidth]{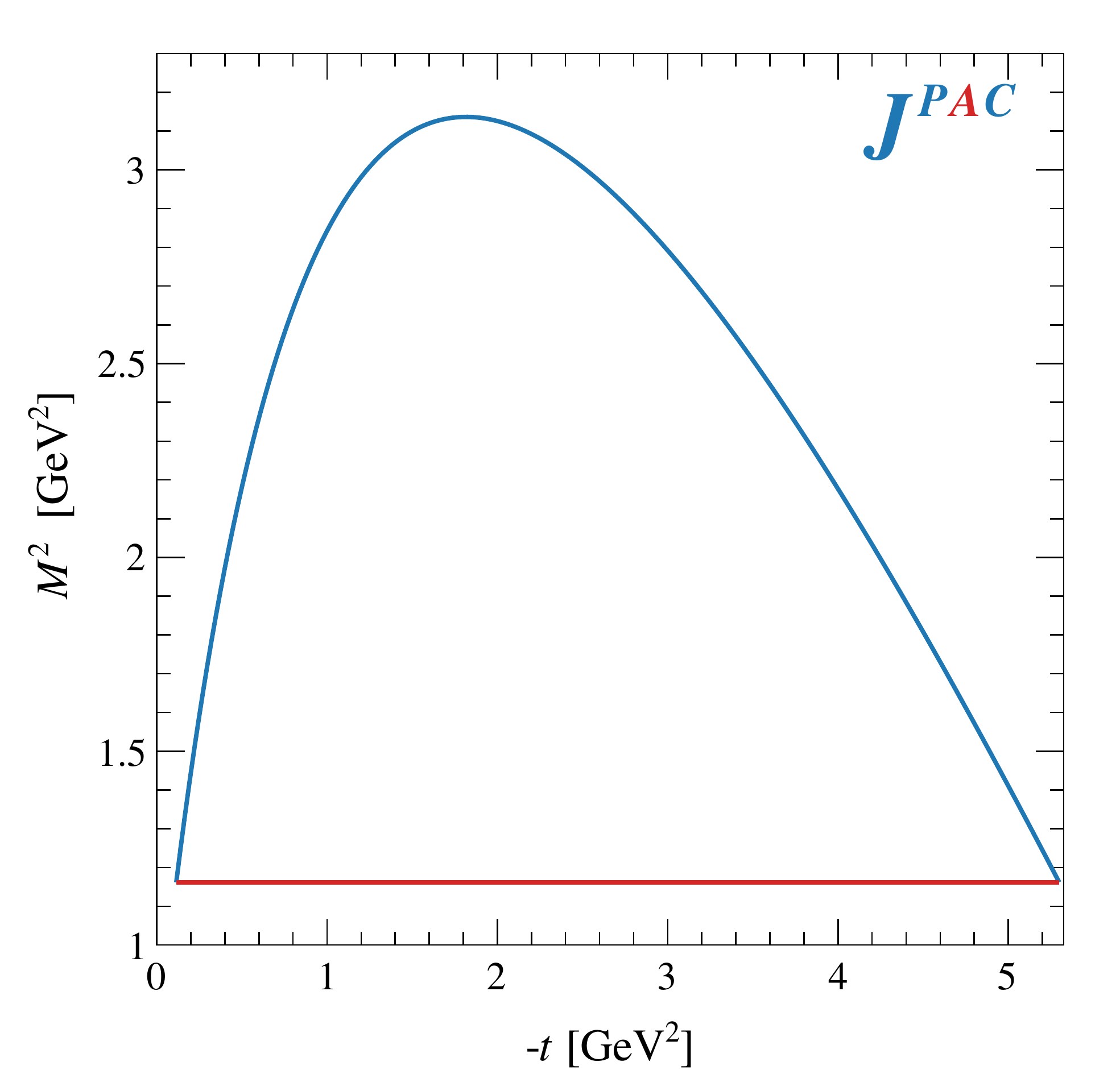}
        \caption{Chew-Low plot marking physical semi-inclusive kinematic region in missing mass and momentum transfer at fixed $W_{\gamma p} = 3~\gev$ for $b_1(1235)$ production. }
        \label{fig:chew_low}
    \end{figure}

We use the standard notations for $2\to2$ scattering, considering the \mN system as a quasi-particle of mass \M. One can thus define the Mandelstam variables $s,t,u$ satisfying $s + t + u = m_p^2 + m_\mQ^2 + \M^2$.
The initial state is fully characterized by the energy of the photon beam in the center-of-mass frame: 
    \begin{equation}
        E_\gamma = q_i = \frac{s - m_p^2}{2\sqrt{s}}, 
    \end{equation}
which only depends on the total invariant mass of the collision, $s = (p+q)^2 = W_{\gamma p}^2$.
The final state can be characterized by different sets of variables. A natural choice for example is the scattering angle and the missing mass, $\left(\theta, \M^2\right)$. 
The allowed region is obviously given by
    \begin{equation}
        \label{eq:boundary_condition_1}
        -1 \le \cos \theta \le 1 
       ~, \quad
        M_\text{min} \le \M \le \sqrt{s} - m_\mQ~.
\end{equation}
Alternatively, one can use a pair of invariants, $\left(t,\M^2\right)$, with $t = k^2 = (q - q')^2$ being the exchanged pion virtuality.
The scattering angle can be expressed in terms of invariants,
\begin{equation}
        \label{eq:costheta}
        \cos\theta = \frac{s (t-u) - m_p^2 \, (m_\mQ^2 - \M^2)}{4 \, s \, q_i \, q_f}~,
\end{equation}
where $q_f$ is the modulus of the 3-momentum of \mQ in the center-of-mass:
    \begin{equation}
        \label{vq}
        q_f = \frac{\lambda^{1/2}(s, m^2_\mQ, \M^2)}{2 \, \sqrt{s}}~,
    \end{equation}
with $\lambda(a,b,c) = a^2 + b^2 + c^2 - 2ab -2ac -2bc$ the standard \Kallen triangular function.
The allowed region can now be given as
\begin{equation}
        \label{eq:boundary_condition}
        \phi(s, t, \M^2) \geq 0 
        \mand 
        \M \geq M_\text{min}~,
\end{equation}
with $\phi$ being the Lorentz-invariant Kibble function~\cite{Kibble:1960zz}:
    \begin{align}
        \phi(s,t,\M^2) &= \left(2 \sqrt{s} \,q_i \, q_f \sin \theta\right)^2 \nonumber \\
        &= stu \nonumber \\
        &+ s  \, m_\mQ^2  (m_p^2 - \M^2) + t \, m_p^2  (m_\mQ^2 - \M^2) \nonumber \\
       &+ m_\mQ^2  m_p^2  (\M^2 - m_p^2 - m_\mQ^2) ~.
    \end{align}
At fixed $\M^2$ the bounds in $t$ may be given in the closed form:
    \begin{align}
        \label{eq:tboundsM2}
        t_{\pm} = m_\mQ^2 - \frac{(s- m_p^2)(s- \M^2 + m_\mQ^2)}{2s} \pm 2 \, q_i \, q_f\,.
    \end{align}
At fixed $t$, the upper bound $\M \leq \sqrt{s} - m_\mQ$ is contained in the boundary condition of \cref{eq:boundary_condition}:
    \begin{align}
        M^2_\text{max} = \frac{\left(s + t - m_p^2- m_\mQ^2\right)  \left(m_p^2 \, m_\mQ^2 - s \, t\right)}{\left(s - m_p^2\right) \left(m_\mQ^2 -t\right)}~,
    \end{align}
as represented in \cref{fig:chew_low}.

Another choice is to replace $t$ with the fraction of 3-momentum carried by \mQ:
    \begin{equation}
        r = \frac{q_f}{\qmax}~.
    \end{equation}
Here, $\qmax = \lambda^{1/2}\!\left(s,m_\mQ^2, \Mmin^2\right)\big/2\sqrt{s} $ is the maximal value of momentum the meson $\mQ$ can have for fixed total energy, which is reached for $\M \to \Mmin$.
The physical kinematic region is greatly simplified in the $\left(r, \cos\theta\right)$-plane to finite square area:
    \begin{equation}
        \label{eq:rcos}
        0 \leq r \leq 1
        \mand
    -1 \leq \cos\theta \leq 1~.
    \end{equation}

\begin{figure}
        \centering
        \includegraphics[width=0.47\textwidth]{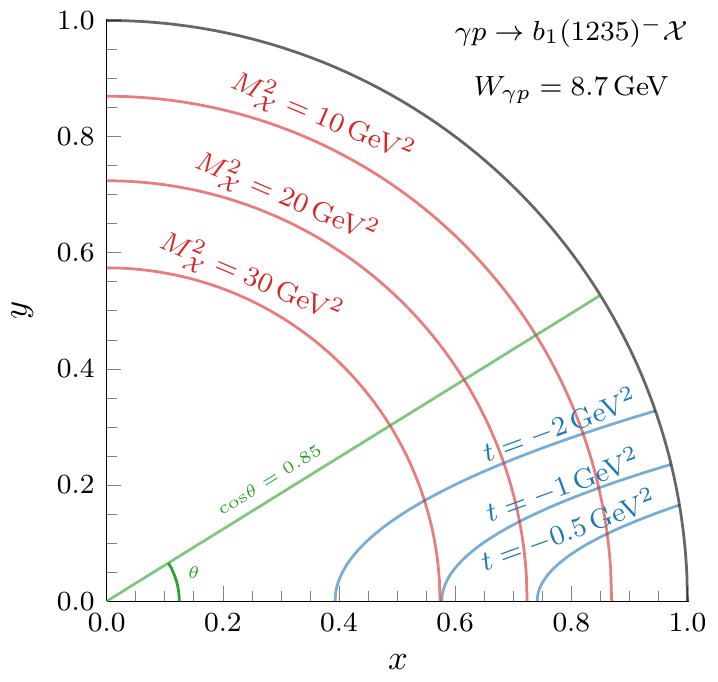}
        \caption{Peyrou plot for $\gamma p \to b_1(1235)^- \mN$ at $W_{\gamma p} = 8.7\gev$. }
        \label{fig:kin_xy}
    \end{figure}

Finally we may consider the Cartesian variables $(x, y)$ defined by:
    \begin{equation}
        \label{eq:xparaperp}
        x = r \, \cos\theta \mand y = r \, \sin\theta ,
    \end{equation}
that correspond to the fractions of longitudinal and transverse 3-momentum, respectively, carried by \mQ.
The allowed region is determined by the circle:
    \begin{equation}
        \label{eq:xy_region}
        x^2 + y^2 \leq 1~,
    \end{equation}
as shown in \cref{fig:kin_xy}.
The Cartesian variables may be related to the polar coordinates by inverting their definition in \cref{eq:xparaperp}, and to the invariant quantities by evaluating
    \begin{equation}
        \label{eq:M2fromEQ}
        \M^2  = s + m_\mQ^2 - 2 \sqrt{s} \, E_\mQ 
       ~,
    \end{equation}
and 
    \begin{equation}
        \label{eq:TfromEQ}
        t = m_\mQ^2 - 2 \, E_\gamma \, E_\mQ + 2 \, q_i \, q_\text{max} \, x~,
    \end{equation}
with the on-shell relation
    \begin{equation}
        \label{eq:EQfromXY}
        E_\mQ^2 = m_\mQ^2 + q^2_\text{max}\, (x^2 + y^2)~.
    \end{equation}

The Lorentz-invariant cross section can be given as:
    \begin{align}\label{DDXS}
        E_\mQ \, \frac{\diff ^3\sigma}{\diff^3q} &= 
        \frac{2\sqrt{s}\, E_\gamma }{\pi} \,\frac{\diff^2\sigma}{\diff t \, \diff \M^2}\nonumber\\
        &= \frac{1}{q^3_\text{max}} \,\frac{E_\mQ}{2\pi \, r^2}  \frac{\diff^2\sigma}{\diff r \, \diff\cos\theta} 
        \nonumber \\
        &= \frac{1}{q^3_\text{max}} \,\frac{E_\mQ}{2\pi \, y}  \frac{\diff^2\sigma}{\diff x \, \diff y} 
       ~.
    \end{align}

For comparison to experimental papers it is convenient as well to look at the mixed variable combination:
    \begin{equation}
        \label{eq:mixedDXS}
        E_\mQ \, \frac{\diff^3\sigma}{\diff^3q} = \frac{1}{\pi} \frac{E_\gamma}{q_\text{max}} \, \frac{\diff^2\sigma}{\diff t \, \diff x}~,
    \end{equation}
although the physical region constraints then become more complicated. For example, at fixed $-1 \leq x \leq 1$ the physical range in $t$ is given by:
    \begin{align}
        \label{eq:tboundsfromX}
        t_{\pm}(x) = 
        m_\mQ^2 &+2 \, q_i \, q_\text{max}\, x \nonumber \\
          &- \frac{(s- m_p^2)(s- M_\pm^2(x) + m_\mQ^2)}{2s},
    \end{align}
where 
\begin{subequations}
    \begin{align}
        \label{eq:m2boundsfromX}
        M^2_+(x) &= \M^2\left(x, y = 0\right)
       ~,
        \\
        M^2_-(x) &= \M^2\left(x, y = \sqrt{1-x^2}\right),
    \end{align}
\end{subequations}
follows from \cref{eq:M2fromEQ,eq:EQfromXY} and refers to the bounds of missing mass at fixed $x$.

For the triple Regge kinematics, where $s$ and $\M^2 \gg |t|$, and $s / \M^2 \gg 1$, we may consider the high energy approximation where $\M^2$ depends on $x$ and $s$ only,
    \begin{equation}
        \label{eq:xinhighe}
        \frac{\M^2}{s} \sim 1-x~,
    \end{equation}
and the upper boundary in $t$ reads    
    \begin{align}
        \label{eq:tapproxboundaries} 
         t_{+}(x) &\sim -m_\mQ^2(1-x) ~.
    \end{align}
Note that this introduces an extra $x$ dependence, while the behavior of $t_-(x)$ is not relevant, as at large $|t|$ the amplitude is exponentially suppressed. 
In conjunction with \cref{eq:mixedDXS} this simplifies numerical calculations. Since Regge forumlae resum leading powers in $\M^2/s$ only, in numerical calculations involving Reggeized exchanges we use the approximation \cref{eq:xinhighe} as to not introduce spurious subleading dependencies on $\M^2/s$. The precise form of \cref{eq:dtsigma_gen} then reads
    \begin{align}
        &\sigma(\gamma p \to \mQ \mN) \sim \int_0^1 \diff x \int_{t_-(x)}^{t_+(x)} \diff t \, \frac{1-x}{16\pi^3}\\
        &\quad\times \left|T_\pi(t) \, \mathcal{P}_\pi\!\left(t,\frac{\M^2}{s} = 1-x\right) \right|^2  \sigmatotst\!\left[\M^2\!\left(s,t,x\right)\right] \nonumber ~.
    \end{align}
The pion-proton cross section is always evaluated at the exact $\M^2$, as using the approximate form would probe the unphysical region below threshold. For the same reason, we use the exact $t_\pm(x)$ from \cref{eq:m2boundsfromX}. This form was used for the curves in \cref{fig:b1_dsigmadx}, while the low-energy cross sections implement the exact kinematics.

\section{$\pi N$ from SAID partial-waves}
\label{app:SAID}
At low energies, we appeal to the amplitudes provided by SAID, which provided partial-wave amplitudes up to orbital angular momentum, $L = 7$ for both $s$-channel isospin projections and parities (\ie 13 waves total) we denote these as $h^{I}_{L\pm}$ where $I = 1/2, \,3/2$ is the $s$-channel isospin projection and $L\pm$ refers to a wave with total spin $J = L\pm 1/2$.

Each partial wave contributes to the total cross section of \cref{eq:sigmaL_pip} with
\begin{align} 
        \sigma_L^{\pi^\pm p}(\M^2) = 
        \frac{\Im \left[C_L^{(+)} \mp C_L^{(-)}\right]}{p_\text{lab}(\M^2) \sqrt{s_0}}
       ~.
    \end{align}
Here the invariant amplitudes $A$, $B$, and $C$ may also be considered at fixed $L$:
    \begin{equation}
        C^{(\pm)}_L = 4\pi \left[A_L^{\pm} + \hat\nu \, B_L^{\pm} \right].
    \end{equation}
Note that these are functions of $\M^2$ only. 
We remind readers that the isospin labels $(\pm)$ refer to $t$-channel isospin projections. As such the amplitude $A$ and $B$ are defined by:
\begin{subequations}
    \begin{align}
        A_L^{\pm} &= \frac{\M+m_p}{E_p+m_p} \, \left(f_L^{\pm} + g_L^{\pm}\right) + \frac{\M-m_p}{E_p-m_p} \, g_L^{\pm}, \\
        B_L^{\pm} &= \frac{1}{E_p+m_p} \, \left(f_L^{\pm} + g_L^{\pm}\right) - \frac{1}{E_p-m_p} \, g_L^{\pm}
       ~,
    \end{align}    
\end{subequations}
and are in terms of the $t$-channel isospin partial waves, $f^\pm_L$ and $g^\pm_L$ and the center-of-mass frame proton energy $E_p = (\M^2 +m_p^2 - m_\pi^2) / (2 \, \M)$. These are constructed from the $s$-channel isospin partial waves (\ie with $I=1/2$ or $3/2$) waves by:
    \begin{subequations}
        \begin{align}
            f_L^{+} &=  \frac{1}{3} \left[f^\frac{1}{2}_{L} + 2 \,  f^\frac{3}{2}_{L} \right] \\ 
            f_L^{-} &=  \frac{1}{3} \left[f^\frac{1}{2}_{L} - f^\frac{3}{2}_{L} \right]
           ~,
        \end{align}
    \end{subequations}
and identical definitions for $g^{\pm}_L$ in terms of $g_L^{\frac{1}{2},\frac{3}{2}}$.

Finally these last $s$-channel partial waves are calculated from the SAID partial wave amplitudes, $h_{L\pm}^{I}$, which are projected also onto definite $J$ by:
    \begin{subequations}
    \begin{align}
        f^{I}_L &= (L+1) \, h^{I}_{L+} + L \, h^{I}_{L-} \\
        \label{eq:gLI}
        g^{I}_L &= \frac{L (L+1)}{2} \left[ h^{I}_{L+} - h^{I}_{L-}\right] 
       ~,
    \end{align}
    \end{subequations}
where the prefactor in \cref{eq:gLI} comes from the value of the first derivative of the Legendre polynomials at $t=0$, \ie $P^\prime_L(\theta = 0)$.

\bibliographystyle{apsrev4-2.bst}
\bibliography{quattro}

\end{document}